\documentclass[aps,amsmath,reprint]{revtex4-1}

\usepackage{hyperref}
\usepackage{graphicx}
\usepackage{bbold}

\hypersetup{colorlinks, linkcolor=blue, citecolor=blue, urlcolor=blue}

\newcommand{\PT}{\mathcal{PT}}
\newcommand{\textPT}{\texorpdfstring{$\PT$}{PT}}

\begin{document}

\title{Connecting active and passive \textPT-symmetric Floquet modulation models}

\author{Andrew K. Harter}
\affiliation{Institute of Industrial Science, The University of Tokyo \\
5-1-5 Kashiwanoha, Kashiwa \\
Chiba 277-8574, Japan}
\author{Yogesh N. Joglekar}
\affiliation{Department of Physics, Indiana University-Purdue University Indianapolis \\ 402 N. Blackford St., Indianapolis, IN 46202, USA}

\date{\today}

\begin{abstract}
Open systems with gain, loss, or both, described by non-Hermitian Hamiltonians, have been a research frontier for the past decade. In particular, such Hamiltonians which possess parity-time ($\PT$) symmetry feature dynamically stable regimes of unbroken symmetry with completely real eigenspectra that are rendered into complex conjugate pairs as the strength of the non-Hermiticity increases. By subjecting a $\PT$-symmetric system to a periodic (Floquet) driving, the regime of dynamical stability can be dramatically affected, leading to a frequency-dependent threshold for the $\PT$-symmetry breaking transition. We present a simple model of a time-dependent $\PT$-symmetric Hamiltonian which smoothly connects the static case, a $\mathcal{PT}$-symmetric Floquet case, and a neutral-$\PT$-symmetric case. We analytically and numerically analyze the $\PT$ phase diagrams in each case, and show that slivers of $\PT$-broken ($\PT$-symmetric) phase extend deep into the nominally low (high) non-Hermiticity region. 
\end{abstract} 
\maketitle

\section{Introduction}
\label{sec:introduction}

Spanning the last two decades, there has been a dramatic increase in the study of systems with dynamics governed by non-Hermitian Hamiltonians. These Hamiltonians lead to behavior with stark differences from their corresponding Hermitian counterparts, including generally complex eigenspectra and non-orthonormal eigenbases; and, as such, they are capable of describing fundamentally open systems which experience external, non-conservative forces arising from the coupling to the surrounding environment.

One important class of non-Hermitian Hamiltonians are those with parity-time ($\PT$) symmetry \cite{Bender1998, Bender2002, Bender2007} which feature spatially separated, balanced sources of gain and loss. In contrast to a general, non-Hermitian case, $\PT$-symmetric Hamiltonians feature a regime of their parameter space, known as the $\PT$-symmetric regime, which is characterized by a completely real eigenspectrum resulting in dynamically stable eigenstates. These eigenvalues remain real until the parameterization of the system reaches an exceptional point  \cite{Heiss2004a, Heiss2004b, Berry2004, Hassan2017} and the $\PT$ symmetry is spontaneously broken, after which, part of the eigenspectrum becomes complex and the dynamics become unstable.

Although first introduced over two decades ago as a complex extension of quantum theory \cite{Bender1998, Bender2002, Mostafazadeh2002, Bender2007},  $\PT$-symmetric Hamiltonians quickly proved their usefulness in describing optical systems \cite{ElGanainy2007, Makris2008} with gain and loss, opening the door for direct experimental observations \cite{Guo2009, Ruter2010}. In the years since, these findings have been extended to a wide variety of experimental setups including waveguide arrays \cite{Szameit2011}, optical resonators \cite{Chang2014, Hodaei2014, Peng2014}, electrical circuits \cite{Schindler2011}, mechanical systems \cite{Bender2013}, acoustics \cite{Fleury2015}, and atomic systems \cite{Zhang2016, Peng2016}. Recently, empowered by state-of-the-art quantum tomography and non-unitary embedding techniques, experimenters have successfully observed $\PT$-symmetric systems at the fully quantum level using advanced photonics \cite{Xiao2017, Tang2016, Klauck2019, Xiao2019, Bian2020}, superconducting qubits \cite{Naghiloo2019}, nitrogen vacancy (NV) centers \cite{Wu2019}, and nuclear magnetic resonance (NMR) quantum computing platforms \cite{Zheng2013, Wen2019}. Consequently, the field has become quite diverse in both theory and experiment. (For further reading see the following review articles \cite{Joglekar2013, Feng2017, ElGanainy2018, Ozdemir2019}).

While the aforementioned studies have largely been focused on static systems, recently, there has been an increased interest in systems which feature time-periodic sources of gain and loss. In these cases, the dynamical stability of the system cannot be determined by analyzing the properties of the time-dependent Hamiltonian at an individual point in time. Specifically, a periodically driven system \cite{Floquet1883, Shirley1965} can be characterized by an effective, time-independent Hamiltonian, called the Floquet Hamiltonian, which accounts for the dynamical effects that occur over an integer number of periods, and the micromotion operator that accounts for the dynamics during one period. These systems feature novel symmetries and topologies \cite{Kitagawa2010, DalLago2015, Fruchart2016} which are experimentally accessible \cite{Rechtsman2013, Wang2013, Jotzu2014, Maczewsky2017, Bordia2017, Wintersperger2020} and, in the non-Hermitian case, can feature unique regimes of stability not found in their static counterparts \cite{Joglekar2014, Lee2015, Gong2015, Longhi2017a, Longhi2017b, Zhou2018, Zhou2019, Li2019b, Zhang2020, Wu2020}.

Previous studies have found that periodic, non-Hermitian Hamiltonians having $\PT$ symmetry exhibit broad regimes of both stability and instability with multiple crossovers occurring at very low frequencies \cite{Joglekar2014, Lee2015}. The stabilizing effect of the periodic modulation leads to new regions of unbroken $\PT$-symmetry and opens the door to exploring frequency-dependent $\PT$-symmetry breaking \cite{Yuce2015a, Yuce2015b, Maamache2017, Turker2018, Harter2020, Li2020, Duan2020, Mochizuki2020}. Furthermore, such phenomena have proved to be amenable to experiment in a variety of settings \cite{Chitsazi2017, Li2019a, LeonMontiel2018}.

In this theoretical study, we present a simple model Hamiltonian which is periodically driven between two $\PT$-symmetric configurations and parameterized such that the model can encompass many of the previously described $\PT$-symmetric systems, connecting them via a single parameter, as well as introducing new regions of interest. We emphasize certain points in the parameterization which correspond to systems which are amenable to various experimental setups. Furthermore, we analyze the interesting regimes of long-term dynamical behavior which arise as a result of carefully choosing the system parameters as well as the type of driving.

The organization of the paper is as follows. In Sec.~\ref{sec:background}, we give a brief review of Floquet driving and the Floquet effective Hamiltonian, connecting its eigenvalues to the dynamical stability of the overall system. In Sec.~\ref{sec:global-model}, we introduce our parameterized, periodic-driving model and explore the resulting $\PT$-phase diagrams associated to various realizations of the model. In Sec.~\ref{sec:analytical-approach}, we give the details of an analytical approach which helps to shed light on some of the frequency dependent results previously discussed. We highlight the specific cases of $\PT$ to reversed-$\PT$-symmetric driving (Sec.~\ref{sec:pt-pt-driving}) and $\PT$ to Hermitian driving (Sec.~\ref{sec:pt-hermitian-driving}). Finally we conclude with a discussion of the overall implications of these results in Sec.~\ref{sec:conclusion}.

\section{Background}
\label{sec:background}

We begin our study with a short description of a prototypical, two-site static $\PT$-symmetric system which provides a generic platform to understand the $\PT$ symmetry breaking phenomenon; this is because, as we will see below, the $\PT$-breaking transition is heralded by level attraction that leads to the exceptional-point degeneracy. We then discuss the theoretical and practical implications of introducing time-dependent driving and its effect on the symmetry breaking phenomena.

For a static Hamiltonian, the $\PT$-symmetry breaking condition is determined by the emergence of complex-conjugate eigenvalues. Consider, for example, the simple $\PT$-symmetric Hamiltonian with gain/loss rate $\gamma$,
\begin{equation}
	H_\PT(\gamma) = \begin{bmatrix} i\gamma & -J \\ -J & -i\gamma \end{bmatrix}=i\gamma\sigma_z-J\sigma_x\,
\label{eqn:ptham}
\end{equation}
where $\sigma_x$ and $\sigma_z$ are standard Pauli matrices. $H_\PT$ commutes with the antilinear $\PT$ operator where $\mathcal{P} \equiv \sigma_x$ and $\mathcal{T} \equiv *$ (complex conjugation), which ensures that the eigenvalues of $H_\PT$ are either purely real or occur in complex conjugate pairs. Indeed the eigenvalues $E_\pm(\gamma) = \pm\sqrt{J^2 - \gamma^2}$ satisfy this. When $\gamma$ is increased from $0$ to $J$, $E_\pm(\gamma)$ both remain real and the non-orthogonal eigenvectors $|\pm(\gamma)\rangle$ of $H_\PT(\gamma)$ are simultaneous eigenvectors of the $\PT$ operator with eigenvalue unity; hence, for $\gamma \leq J$, $H_\PT(\gamma)$ has unbroken $\PT$-symmetry. It is easy to see that when $\gamma > J$, the eigenvalues $E_\pm$ are pure imaginary, complex conjugates. Due to the  antilinearity of the $\PT$ operator, however, the eigenvectors obey $\PT|+\rangle=|-\rangle$, meaning the $\PT$ symmetry is broken. We note that the (Dirac) inner product of the two eigenstates is given by $|\langle+|-\rangle|=\min(\gamma/J,J/\gamma)\leq 1$, and thus reaches unity at the exceptional point $\gamma=J$. 

A system such as that in Eq.(\ref{eqn:ptham}) is known as an active $\PT$-symmetric system. However, many properties of the $\mathcal{PT}$-transition, including the presence of an exceptional point, are also shared by Hamiltonians with only mode-selective losses. The latter are called passive $\PT$ systems, and they can be derived from active $\PT$ systems by shifting the reference energy level by an imaginary amount $-i\gamma$.

Thus, a prototypical lossy (or passive) $\PT$ Hamiltonian is given by $H_L=H_\PT-i\gamma{\mathbb{1}}_2$~\cite{Guo2009}. Note that the eigenvalues of $H_L$ are always complex and are indicative of non-orthogonal eigenmodes that decay with time. At small $\gamma$, the decay rates of the two modes are identical. With certain abuse of terminology, this region is referred to as a ``$\PT$ symmetric region''. Beyond a critical loss strength, the two decay rates become different and the system enters a ``$\PT$ symmetry broken region''. We note that this definition relies on the lossy Hamiltonian $H_L$ being identity-shifted from a genuine $\PT$  symmetric Hamiltonian. Alternatively, one can define the $\PT$-breaking transition as the emergence of a ``slowly decaying'' eigenmode, whose lifetime increases with increasing $\gamma$. The latter criterion has been experimentally used to characterize the passive $\PT$ transition by loss-induced transparency. However, it is important to note that the emergence of a slowly decaying eigenmode does not depend upon the existence of an exceptional point~\cite{LeonMontiel2018, Joglekar2018}.

Consequentially, mode-selective-lossy or passive systems, while not themselves being $\PT$-symmetric, drastically increase the range of experimental setups which are possible, as the stringent requirement for matched gain and loss is reduced to a necessity for pure loss only. They also permit us to extend the ideas of $\PT$ symmetry and exceptional points into the truly quantum domain. Due to the quantum limits on noise in linear amplifiers~\cite{Caves1982}, an active $\PT$-system, i.e. a system with balanced gain and loss is not possible~\cite{Scheel2018}; however, a passive $\PT$-system can be realized by appropriately post-selecting over the quantum trajectories of a Lindblad evolution~\cite{Naghiloo2019}. 

In contrast to the static case where $\PT$ symmetry breaking occurs when the non-Hermitian strength is equal to the Hermitian one, the dynamics are far richer for a minimal model with periodic time dependence. For example, if we let the gain/loss strength $\gamma$ from Eq.~(\ref{eqn:ptham}) be driven by some function, $\gamma(t)$, the eigenvalues of the instantaneous Hamiltonian $H_\PT(t)$ can change between real and complex conjugates depending on time and the functional form of $\gamma(t)$. Since the time-translational invariance is, in general broken, no statement can be made about whether the system is in the $\mathcal{PT}$-symmetric phase or $\PT$-broken phase. However, for a time-periodic Hamiltonian with period $T$, i.e. $H_\PT(t+T)=H_\PT(t)$, according to the Floquet theorem \cite{Floquet1883, Shirley1965}, the long-term dynamics of the system are captured by the time-evolution operator over one period ($\hbar=1$)
\begin{equation}
	G(T) = \mathrm{T} e^{-i\int_{0}^T dt' H_\PT(t')}.
\end{equation}
Here, $\mathrm{T}$ stands for the time-ordered product that takes into account the non-commuting nature of Hamiltonians at different times, $[H_\PT(t),H_\PT(t')]\neq 0$. The non-unitary $G(T)=\exp(-iH_FT)$, in turn defines the effective, non-Hermitian Floquet Hamiltonian $H_{F}(J,\gamma,\omega=2\pi/T)$ which encapsulates the average effects of the periodic driving. Thus, by analyzing eigenvalues of $G(T)$ or, equivalently the quasienergies $\varepsilon_F$ of the Floquet Hamiltonian $H_F$, we are able to determine the long-term behavior of the system, including whether the system is in the $\PT$-symmetric phase (purely real quasienergies) or $\PT$-broken phase (complex conjugate quasienergies).

\section{Global Model}
\label{sec:global-model}

\begin{figure}
	\centering
	\includegraphics[width=\columnwidth]{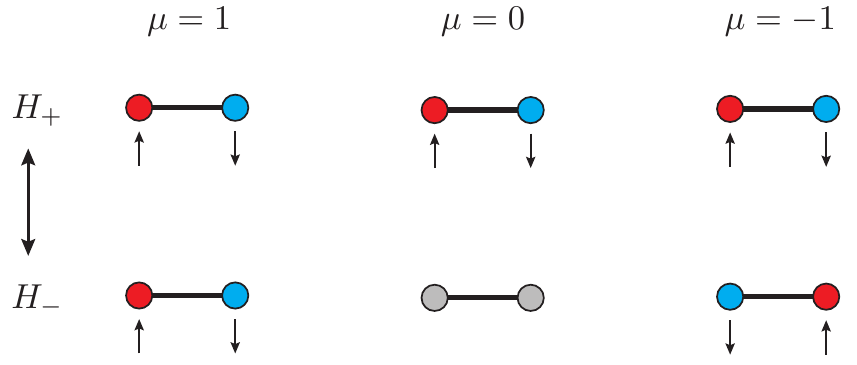}
	\caption{$\PT$-symmetric Floquet modulation given by Eq.~(\ref{eqn:floq-driving}) and parameterized by $\mu=(1-\delta)/(1+\delta)$. Each column shows a pair of diagrams depicting the system governed by $H_+$ (top row) and $H_-$ (bottom row) respectively. Two sites with coupling $J$ are colored according to whether a site has a gain $+i\gamma_0$ (red), a loss $-i\gamma_0$ (blue) or is neutral (gray). The first column, with $\mu = 1$, represents the static case, $H_{+}=H_{-}$. The central column, with $\mu = 0$, corresponds to a system evolving with a $\PT$-symmetric Hamiltonian for half the period and with a Hermitian Hamiltonian $-J\sigma_x$ for the other half. The third column, with $\mu = -1$, shows a $\PT$-symmetric system in which gain and loss positions switch after half the period, i.e. $H_{-}=H_{+}^*=\mathcal{P}H_{+}\mathcal{P}$. The three columns are characterized by increasing ratio of gain-loss variation to its mean value.}
	\label{fig:mu-driving}
\end{figure}

In this section we describe a $\PT$-symmetric Floquet Hamiltonian with a parameter space which encapsulates both passive and active $\PT$-symmetric models as well as Hermitian ones. Consider the general two-step driving between two static Hamiltonians $H_+$ and $H_-$, where $H_\pm \equiv H_\PT(\gamma_\pm)=-J\sigma_x+i\gamma_\pm\sigma_z$ and $\sigma_x,\sigma_z$ are standard Pauli matrices \cite{Li2019a, Duan2020}. We characterize the two non-Hermitian strengths as $\gamma_\pm \equiv \bar{\gamma}(1 \pm \delta)$ where, without loss of generality $\delta\geq 0$. Thus, $\bar{\gamma}>0$ defines the  average gain-loss strength, and $\delta$ is the fractional deviation from the average in each step of the driving (see the summary in Fig.~\ref{fig:mu-driving}). Thus, the system is governed by a time-periodic, piecewise constant Hamiltonian
\begin{equation}
	H(t) = \begin{cases}
		H_+ \,,\quad 0 \leq t < T/2 \\
		H_- \,,\quad T/2 \leq t < T
	\end{cases}
	\label{eqn:floq-driving}
\end{equation}
in one period $T$, so that $H$ equals $H_+$ up to time $T/2$ where it is abruptly changed to $H_-$ for the rest of the period. 

Consequently, the time evolution operator over one period is merely the product of two static time evolution operators
\begin{equation}
	G(T) = e^{-iH_-T/2}e^{-iH_+T/2}=e^{-iH_FT} \,.
\end{equation}
where we have set $\hbar = 1$ and the non-trivial effects in the Floquet Hamiltonian $H_F$ arise because $[H_{+},H_{-}]\neq 0$. When $\delta=0$, we obtain the static case, with a single threshold at $\bar{\gamma}=J$ irrespective of the period $T$, or equivalently the frequency $\omega=2\pi/T$. As $\delta$ is increased from zero to unity, $\gamma_-\rightarrow 0$, so that $H_+$ is a $\PT$-symmetric Hamiltonian and $H_-$ is a Hermitian one. Finally, when $\delta \gg 1$, $\gamma_\pm \rightarrow \pm\delta\bar{\gamma}$ and the periodic modulation is between $\PT$-symmetric Hamiltonians with reversed gain and loss locations.

Motivated by this parametrization, we can set $\gamma_+/J= \bar{\gamma}/J_0$ and held constant with respect to $\delta$, so that as $\delta \rightarrow \infty$, $J_0$ sets a fixed coupling scale and the Hamiltonians scale accordingly. If we define the ratio of the gain-loss strengths in each half of the period
\begin{equation}
	\mu = \frac{\gamma_{-}}{\gamma_{+}}=\frac{1 - \delta}{1 + \delta} \,,
	\label{eqn:mu}
\end{equation}
so that $\delta=0$ gives $\mu=1$, $\delta=1$ gives $\mu=0$, and $\delta \rightarrow \infty$ gives $\mu\rightarrow -1$ (Fig.~\ref{fig:mu-driving}), we see that $H_+ = H_\PT(\gamma_0)$ and $H_- = H_\PT(\mu\gamma_0)$ where $\gamma_0\equiv \bar{\gamma}(J/J_0)$. At this point the connection is clear, and we may focus on analyzing the $\PT$-symmetry breaking properties of such time-periodic, non-Hermitian Hamiltonians that modulate between $H_+ \equiv H_\PT(\gamma_0)$ and $H_- \equiv H_\PT(\mu\gamma_0)$, where the driving is parameterized by the choice of  $\mu$ along with the scaled frequency $\omega/J$, and the relative gain and loss strength is parameterized by $\gamma_0/J$.

For an active $\PT$-symmetric system, the meaning of this driving is clear, given one has sufficient control over the sources of gain and loss in the system. However, for a passive $\PT$-symmetric system, there is no gain. In this case, the results for the passive system can be obtained by translating the problem from an active $\PT$-symmetric system by a simple shift of the Hamiltonians $H_+$ and $H_-$ by a negative imaginary amount proportional to the identity, $-i\gamma_0\mathbb{1}_2$ and $-i|\mu|\gamma_0\mathbb{1}_2$ respectively, which is nothing but an overall loss factor $e^{-(|\mu|+1)\gamma_0T/2}$ in the time evolution over one period. 

Passive systems, as discussed in the last section, present an advantage easily seen in the static case, where only a single site must have loss and the other can be kept neutral. Now, for $\gamma_0 > 0$, in a passive system, this corresponds to alternating the rate of loss on a single site. Similarly, for the case of $\mu = 0$, the passive case simply corresponds to on/off pulsing of the loss in the system. However, when $\mu < 0$ in a passive system, the situation can no longer be achieved by controlling the loss of just one of the sites; rather precise control over the rate of loss on both sites must be obtained to implement these cases.

\begin{figure*}
	\centering
	\includegraphics[width=\textwidth]{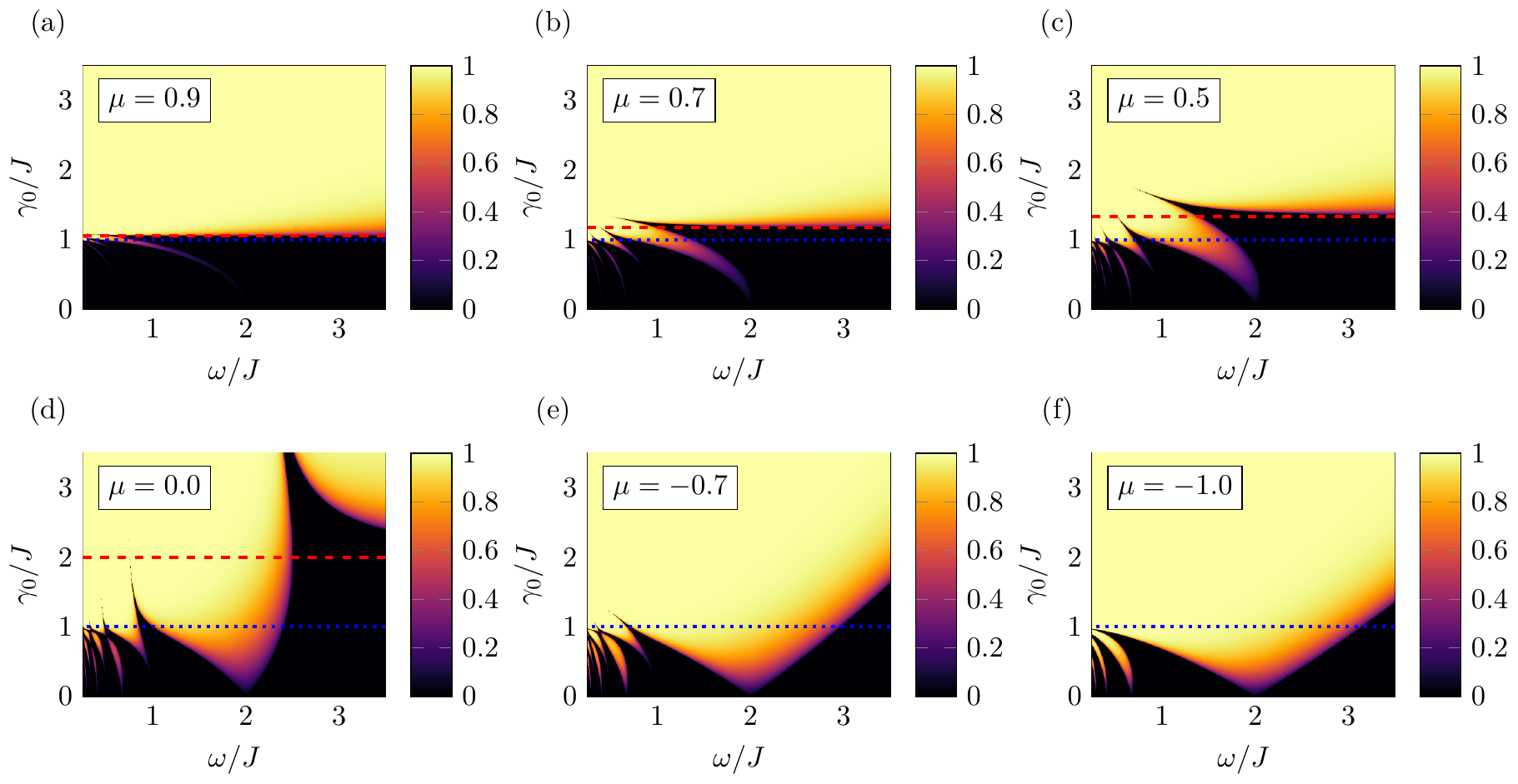}
	\caption{$\PT$ phase diagram of the Floquet driving model in Eq.~(\ref{eqn:floq-driving}) with modulation parameter $\mu$ (see Eq.~(\ref{eqn:mu})) over the parameter space spanned by $\gamma_0/J$ and $\omega/J$, colorized by $c$ defined in Eq.~(\ref{eqn:ccolor}). We also indicate the static $\PT$-symmetry breaking threshold ($\gamma_0/J=1$) and the high-frequency effective $\PT$-symmetry breaking threshold $(\gamma_\infty)$ by the dotted blue and dashed red lines respectively. In (a), $\mu = 0.9$, which shows a $\PT$ phase diagram very close to the static situation; we note that a small sliver of the static $\PT$-broken phase begins to extend down into the region which is unbroken in the static case. In (b), $\mu = 0.7$, and the high frequency $\PT$-breaking threshold has clearly increased and separated from the static threshold, and the $\PT$-broken regime clearly extends all the way to $\gamma_0 = 0$ at the primary resonance frequency $\omega/J = 2$. In (c), $\mu = 0.5$, and additional slivers of the broken phase extend down below $\gamma_0/J = 1$; further, we clearly see regions of $\PT$-unbroken phase extend above $\gamma_0/J = 1$. In (d), $\mu = 0$, which corresponds to driving between a $\PT$-symmetric Hamiltonian and a Hermitian one. In this case, we clearly see many slivers of the $\PT$-broken phase have extended down to near $\gamma_0 = 0$ at specific resonance frequencies which are located at $\omega/J = 2/n$ for integers $n$. In (e), $\mu = -0.7$, and we note the recession of the $\PT$ broken phase in the regions which had previously extended down to the even-numbered resonance frequencies at $\omega/J = 2/n$ for even $n$. Similarly, the regions of $\PT$-unbroken symmetry reaching above $\gamma_0/J = 1$ have receded as well. In (f), $\mu = -1$, corresponding to driving between two exactly reversed $\PT$-symmetric Hamiltonians. In this case, the regions of extended $\PT$-broken phase corresponding to even-numbered resonance frequencies have completely disappeared, as have the regions of unbroken $\PT$ symmetry above $\gamma_0/J = 1$.}
\label{fig:ptphase-mu}
\end{figure*}

In general, when the frequency of periodic modulation is sufficiently high ($\omega$ is much larger than the magnitude of the maximum eigenvalue of $H_+$ and $H_-$), the Floquet Hamiltonian takes on an approximate form which is the average of $H_+$ and $H_-$, namely, $H_F \rightarrow H_\PT((1+\mu)\gamma_0/2)$ as $\omega/\gamma_0\rightarrow\infty$, and the $\PT$-symmetry breaking condition approaches an effective static $\PT$-symmetry breaking threshold
\begin{equation}
	\gamma_\infty = \frac{2J}{|1+\mu|}\,.
\label{eqn:high-freq-thresh}
\end{equation}

In Fig.~\ref{fig:ptphase-mu}, we show the numerically obtained $\PT$-phase diagram for the $(\gamma_0/J,\omega/J)$ parameter space at six different values of $\mu=\{0.9, 0.7, 0.5, 0, -0.7, -1\}$ in panels (a)-(f) respectively. These values provide snapshots indicating the changes to the $\PT$-phase diagram over the same section of the parameter space spanned by $\gamma_0$ and $\omega$. The color shown in each panel indicates the normalized amplification rate, i.e. 
\begin{equation}
	c \equiv \frac{|g_+| - |g_-|}{|g_+| + |g_-|} \,,
\label{eqn:ccolor}
\end{equation}
where $g_\pm$ are the eigenvalues of the time evolution matrix after one period, $G(T)$, arranged such that $|g_+| \geq |g_-|$. Thus, $c$, is a measure of the $\PT$-symmetry breaking; specifically, when $H_F$ has all real eigenvalues, $c = 0$. In each panel, the dotted, blue horizontal line indicates the static ($\mu=1$) $\PT$-symmetry breaking threshold, which is independent of $\omega$; further, to contrast the changes in each case, the high frequency effective static $\PT$-threshold $\gamma_\infty$ is shown by the dashed, red horizontal line. 

In Fig.~\ref{fig:ptphase-mu}(a)-(c), $\mu > 0$, so that in the high frequency limit, the effective static $\PT$-breaking threshold is $J < \gamma_\infty < 2J$ according to Eq.~(\ref{eqn:high-freq-thresh}). In (a), we show the $\PT$ phase diagram for $\mu = 0.9$. This case is only slightly removed from the static situation, and, as expected, the phase diagram is largely independent of $\omega$; however, a small region of the $\PT$-symmetry broken phase has begun to extend down into the region in which the $\PT$ symmetry is unbroken in the static case. As $\mu$ is reduced to $0.7$ in (b), this region of extended $\PT$-broken phase continues to reach down until it is clear that it approaches the primary resonance frequency of $\omega_0/J = 2$ near $\gamma_0 = 0$. As we decrease $\mu$ further to $0.5$, as in (c), we see this region expanding as well as the introduction of many other similar regions of broken $\PT$ symmetry below the static $\PT$ threshold at $\gamma_0 /J= 1$. Furthermore, we see that the region of unbroken $\PT$ symmetry has begun to extend above $\gamma_0/J = 1$ even in the lower driving frequency regime ($\omega/J < 2$), and in the high frequency regime, we see that the effective $\PT$ threshold has, at this point, increased significantly from the true static case of $1$ to $\gamma_\infty/J = 4/3$.

In Fig.~\ref{fig:ptphase-mu}(d), the $\PT$ phase diagram is shown for the case where $\mu = 0$, which corresponds to driving between a $\PT$-symmetric Hamiltonian and a Hermitian one. In this case, previous experiments have successfully probed the frequency-dependent crossovers between broken and unbroken $\PT$ symmetry which exist in this system \cite{Li2019a, LeonMontiel2018}. Here, we see the increased size of the $\PT$-symmetry breaking regions which  began forming in (a)-(c), and we note that they extend down to small, but nonzero, values of $\gamma_0$ in the vicinity of $\omega/J = 1/n$ for integers $n$.  Likewise, the regions of unbroken $\PT$ symmetry now extend deep into the portion of the parameter space where $\gamma_0/J \geq 1$. By contrast, in the high driving frequency regime, the non-Hermitian portions of $H_+$ and $H_-$ do not fully cancel leading to $\PT$ symmetry breaking for $\gamma_0/J \geq 2$ in this region, and the lack of stable topological states \cite{Harter2020}.

Next, for both Fig.~\ref{fig:ptphase-mu}(e) and (f), $\mu < 0$, and we see the high frequency effective static threshold begin to increase far above $J$. In (e), we show the case for $\mu = -0.7$. Here, we observe the reduction of the regions of $\PT$-symmetry breaking which had previously extended down  to small values of $\gamma_0$ approaching the resonance frequencies $\omega/J = 2/n$ specifically for even values of $n$. A slight further decrease of $\mu$ to $-1$ gives the case shown in (f), which corresponds to periodic driving between two opposite $\PT$-symmetric systems with comparatively reversed gain and loss. This type of driving leads to specific resonant frequencies $\omega/J = 2/n$ only for odd integers $n$; in the neighborhood of these frequencies, even for small gain/loss $\gamma_0$, the system remains in the $\PT$-broken phase. Furthermore, below a critical driving frequency $\omega$, the $\PT$ symmetry is always broken for $\gamma_0/J \geq 1$. Importantly, in the high driving frequency regime, where $\omega \gg \gamma_0$, this system appears Hermitian because the averaging effect between $H_+$ and $H_-$ leads to the cancellation of the gains and losses. For this configuration, this system has been shown to support dynamically stable topological states which survive precisely due to this type of cancellation \cite{Yuce2015b, Harter2020}.

Qualitatively, these results show us that, starting from a purely static situation at $\mu = 1$, with a $\PT$-symmetry breaking threshold at $\gamma_0/J = 1$, as we decrease $\mu$ to zero, many slivers of broken $\PT$ symmetry extend down into the statically unbroken $\PT$ phase, eventually bringing the broken phase down near $\gamma_0 = 0$ in the neighborhood of resonance frequencies $\omega/J = 2/n$ for integers $n$. In this same traversal over $\mu$, we also see the introduction of similar regions of unbroken $\PT$ symmetry extending into the statically broken regime. However as $\mu$ is decreased further from $0$ to $-1$, we see that these extensions of the unbroken phase into the broken phase are removed, and the extensions of the broken phase into the unbroken phase which correspond to resonance frequencies associated with even integers $n$ are also removed. In the next section we will analytically compare the two cases of $\mu = 0$ and $\mu = -1$ to provide insight into this situation.

\section{Analytical Approach}
\label{sec:analytical-approach}

To gain a better understanding of the numerical results presented in Sec.~\ref{sec:global-model}, in this section, we approach the problem analytically and determine the form of the Floquet effective Hamiltonian $H_F(\gamma/J,\omega/J)$. By analyzing the quasienergies of the effective Hamiltonian, we can locate the regions of broken and unbroken $\PT$ symmetry,  and thereby understand the phase diagram differences for different values of $\mu$. 

We begin by finding the time evolution up to the first period ($t = T$) by manually multiplying out the product $G(T) = G_-(T/2)G_+(T/2)$. Here $G(T) = \exp(-iH_FT)$ defines the effective Floquet Hamiltonian $H_F$, and $G_\pm(t) = \exp(-iH_\pm t)$ are the time evolution operators associated to the static Hamiltonians $H_\pm$. 

We note that, after the scaling, the periodic modulation switches the system between two Hamiltonians $H_{1,2} = \vec{r}_{1,2} \cdot \vec{\sigma}$, where we define the complex vectors $\vec{r}_1 = (-J, 0, i\gamma_0)$ and $\vec{r}_2 = (-J, 0, i\mu\gamma_0)$, and $\vec{\sigma}=(\sigma_x,\sigma_y,\sigma_z)$ is the usual vector of Pauli matrices. Then, the evolution up to one period $T=2\tau$ is defined by $G(2\tau) = G_2(\tau)G_1(\tau)$, where each time evolution operator $G_k(\tau)$ can be written as 
\begin{align}
	G_k(\tau) &= \cos( r_k\tau){\mathbb{1}}_2 - i \sin(r_k\tau) (\hat{r}_k \cdot \vec{\sigma}) \,
\end{align}
where $r_1 = \sqrt{J^2 - \gamma_0^2}$ and $r_2 = \sqrt{J^2 - \mu^2\gamma_0^2}$ are eigenvalues of the static Hamiltonians $H_{1,2}$. 
The resulting time-evolution operator $G(T)$, a non-unitary, invertible $2\times2$ matrix, can be represented as a linear combination of the identity and three Pauli matrices, i.e.
\begin{align}
	& \nonumber G(2\tau)  = \left[\cos(r_2\tau) \cos(r_1\tau)-\sin(r_2\tau)\sin(r_1\tau) (\hat{r}_2\cdot\hat{r}_1)\right]\mathbb{1}_2\\
	\nonumber & - i[\cos(r_2\tau)\sin(r_1\tau)\hat{r}_1+ \sin(r_2\tau)\cos(r_1\tau)\hat{r}_2 \\
	& + \sin(r_2\tau)\sin(r_1\tau) (\hat{r}_2 \times \hat{r}_1)]\cdot\vec{\sigma} \,.
\end{align}
On the other hand, by parametrizing the Floquet Hamiltonian in terms of the magnitude and direction of an effective field, i.e. $H_F=\varepsilon_F \hat{r}_F\cdot\vec{\sigma}$, it follows that $G(2\tau)=\cos(2\varepsilon_F\tau)\mathbb{1}_2-i\sin(2\varepsilon_F\tau)(\hat{r}_F\cdot\vec{\sigma})$. By equating the coefficients of Pauli matrices, we 
obtain 
\begin{align}
	\begin{split}
		\cos(2\varepsilon_F\tau) &=  \cos(r_2\tau)\cos(r_1\tau) \\
		&-\frac{J^2- \mu\gamma_0^2}{r_1r_2}\sin(r_2\tau)\sin(r_1\tau) \,,
	\end{split}
	\label{eqn:floquet-cos}\\
	\sin(2\varepsilon_F\tau)\hat{r}_F &= a_x\hat{x} + ia_y\hat{y} + ia_z\hat{z} \,,
	\label{eqn:floquet-sin}
\end{align}
where the dimensionless components of the effective field direction $\hat{r}_F$ are given by 
\begin{align}
	a_x &= \frac{J}{r_1}\cos(r_2\tau)\sin(r_1\tau) +\frac{J}{r_2}\sin(r_2\tau) \cos(r_1\tau), \nonumber\\
	a_y &= (\mu-1)\frac{J\gamma_0}{r_1r_2}\sin(r_2\tau)\sin(r_1\tau),\nonumber\\
	a_z &= \frac{\gamma_0}{r_1}\cos(r_2\tau)\sin(r_1\tau)+\mu\frac{\gamma_0}{r_2}\sin(r_2\tau)\cos(r_1\tau).\nonumber
\end{align}

We remind the reader that because $r_1$ and $r_2$ are either real or pure imaginary, $a_x$, $a_y$, and $a_z$ are all real quantities. Also, since $r_2$ is even in $\mu$, at $\mu = \pm 1$, $r_2 = r_1$. Thus, when $\mu = -1$, which corresponds to $\PT$ to reversed-$\PT$ driving, $a_z = 0$, and the effective Floquet Hamiltonian has the symmetry $\sigma_z H_F \sigma_z = -H_F$. Similarly, when $\mu=1$ (the static case), $a_y = 0$, and the Floquet Hamiltonian reduces to the expected static one.

Note that the eigenvalues of the effective Floquet Hamiltonian $\pm\varepsilon_F$ are determined by Eq.~(\ref{eqn:floquet-cos}). In the following subsections, we examine two special cases with $\mu=\{-1, 0\}$ ($\mu = 1$ is the trivial, static case).

\subsection{{\textPT} to reversed-{\textPT} driving: $\mu=-1$}
\label{sec:pt-pt-driving}

When $\mu = -1$, the system switches between two $\PT$-symmetric Hamiltonians with opposite locations of gain and loss $\gamma_0$ which are equally matched in magnitude. The general $\PT$-phase diagram for this situation is depicted in Fig.~\ref{fig:ptphase-mu}(f). In this case, since $|\mu| = 1$, $r_1 = r_2$, and we have
\begin{equation}
	\cos(2\varepsilon_F\tau) = \cos^2(r_1\tau) - \frac{J^2 + \gamma_0^2}{J^2 - \gamma_0^2}\sin^2(r_1\tau). 
\end{equation}
The $\PT$ symmetric region is determined by the requirement $\cos(2\varepsilon_F\tau)\leq 1$, which simplifies to $\left\lvert\sin(r_1\tau)\right\rvert\leq |r_1|/J$. In this form, we see that when $r_1\tau = n\pi$, which corresponds to an ellipse 
\begin{equation}
\label{eqn:ellipse}
	\gamma_0^2 + n^2\omega^2 = J^2 \,,
\end{equation}
the two eigenvalues $\pm\varepsilon_F$ are real indicating a $\PT$-symmetric phase. On the other hand, when $r_1\tau=(n + 1/2)\pi$ or equivalently, $\gamma_0^2 + (n+1/2)^2\omega^2 = J^2$,  the system has purely imaginary eigenvalues for all $\gamma_0\geq0$ indicating a $\PT$-symmetry broken phase. 

\begin{figure*}
	\centering
	\includegraphics[width=\textwidth]{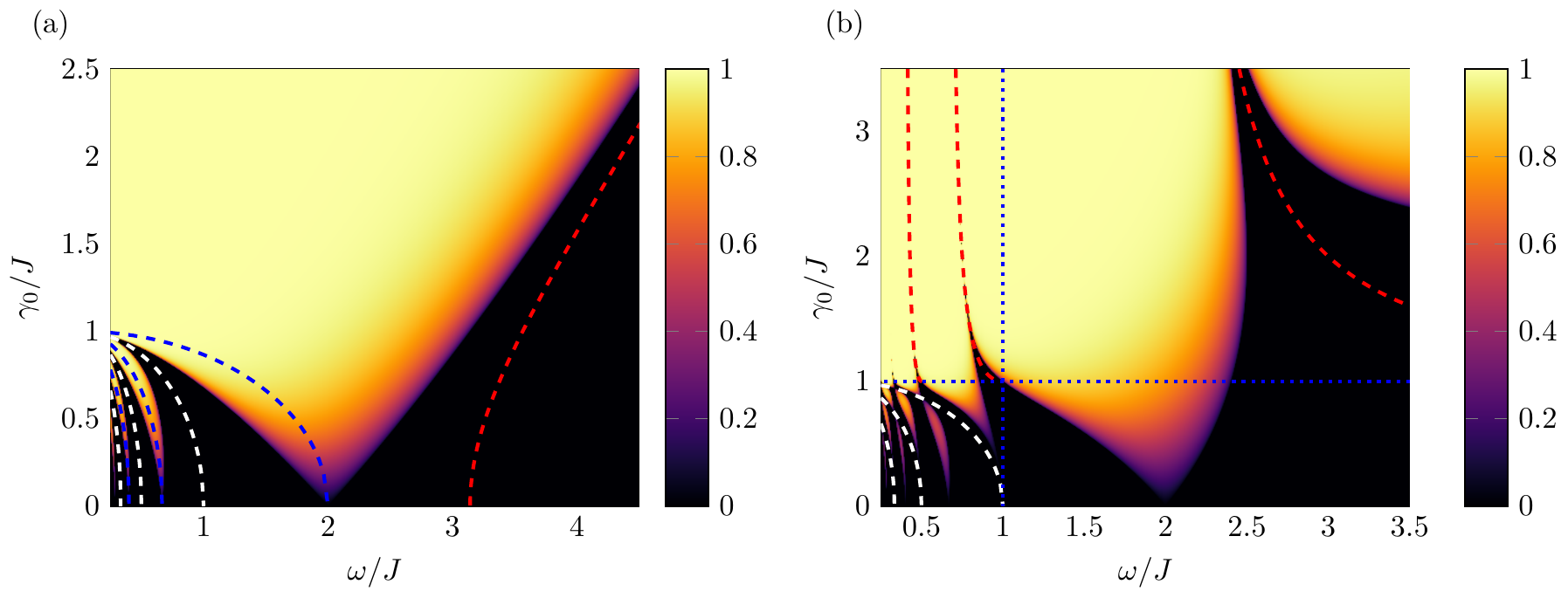}
	\caption{Detailed comparison of the $\PT$ phase diagrams for $\mu = -1$ in (a) and $\mu = 0$ in (b), which correspond to the the systems of Fig.~\ref{fig:ptphase-mu}(f) and (d) respectively. In (a) we show dashed lines of $\PT$-broken (blue) and $\PT$-unbroken (white) symmetry. The blue lines follow the $\PT$-broken phase along lines which, for small $\gamma_0$, extend down to $\omega/J = 2/n$ with $n$ chosen as odd integers, and the white lines follow the $\PT$-unbroken phase along lines which, for small $\gamma_0$ correspond exactly to $\omega/J = 2/n$ for even integers. We also indicate by a red dashed line the approximate $\PT$ phase boundary of Eqn.~(\ref{eqn:pt-boundary}), valid for large values of $\gamma_0$ and $\omega$. In (b), we show dashed lines of $\PT$-unbroken symmetry for values of $\gamma_0$ below the static $\PT$ threshold (white) and also above it (red). The white lines follow the same trajectories as in (a); however, the red lines follow a path described in Eq.~(\ref{eqn:resinfty}) which, for $\gamma_0 \gg J$, approaches the constant frequency value $\omega/J = 2/n$ with odd integers $n$. A horizontal dotted blue line indicates the static $\PT$ symmetry breaking threshold, and a vertical, dotted blue line indicates the constant frequency value $\omega/J = 1$.}
	\label{fig:ptphase-detail}
\end{figure*}

In Fig.~\ref{fig:ptphase-detail}(a), we show the details of the $\PT$ phase diagram arising from such an analysis for the effective Floquet Hamiltonian in the $(\gamma_0/J,\omega/J)$ plane. We have plotted these elliptical sections representing $\PT$ symmetric regions (dashed white line) and $\PT$-broken region (dashed blue line) to highlight their position in the parameter space. We see that the $\PT$-broken phase extends in the low gain/loss limit ($\gamma_0/J \ll 1$) down to the resonance frequencies at $\omega/J = 2/n$ for odd integers $n$. We also see that the $\PT$-unbroken phase for the low driving frequency regime $\omega/J \leq 2$ does not extend above $\gamma_0/J = 1$. 

Furthermore, the compact criterion $|\sin(r_1\tau)|=|r_1|/J$ also allows us to obtain the boundary between the two phases at large values of $\gamma_0$ and $\omega$. In the high-frequency regime defined by $\omega/J \gg 1$ and $\gamma_0/J \gg 1$, the static eigenvalues approach $r_1\approx i\gamma$. This leads to an approximate phase boundary in this regime defined by
\begin{equation}
	\omega = \frac{\pi\gamma}{\sinh^{-1}(\gamma/J)} \approx  \frac{\pi\gamma}{\log(2\gamma/J)} \,,
\label{eqn:pt-boundary}
\end{equation}
and the $\PT$-broken phase boundary peels back with a slowly steepening slope as $\omega$ is increased, paving the way for the completely Hermitian result $H_F = -J\sigma_x$ in the very high frequency limit with $\omega \gg \gamma$, as expected.

\subsection{{\textPT} to Hermitian driving: $\mu=0$}
\label{sec:pt-hermitian-driving}

Another interesting point to examine is the situation for $\mu = 0$, i.e. when the Hamiltonian is Hermitian for half the period. For this configuration, the general $\PT$ phase diagram is depicted in Fig.~\ref{fig:ptphase-mu}(d). In this case, $r_2 = J$ so we have
\begin{equation}
	\cos(2\varepsilon_F\tau) = \cos(r_1\tau)\cos(J\tau) - \frac{J}{r_1}\sin(r_1\tau)\sin(J\tau) \,.
\label{eqn:pt-h-cos}
\end{equation}
We can see that when $r_1\tau=r_1\pi/\omega=n\pi$, we obtain $\cos(2\varepsilon_F\tau) = \pm\cos(J\tau)$ thus ensuring real Floquet quasienergies $\varepsilon_F$ and thus a $\PT$-symmetric phase along the ellipse of Eq.~(\ref{eqn:ellipse}). 

Importantly, with this type of driving, we can also show that the $\PT$-unbroken phase extends deep into the large $\gamma_0/J\gg 1$ region. In this regime, we can approximate $r_1 = iq=i\sqrt{\gamma_0^2-J^2}$ and 
\begin{equation}
	\cos(2\varepsilon_F\tau) \approx\frac{e^{q\tau}}{2}\left[\cos(J\tau) -\frac{J}{q}\sin( J\tau) \right]. 
\end{equation}
By requiring $\cos(2\varepsilon_F\tau) \rightarrow 0$ or equivalently, 
\begin{equation}
	\cot J\tau = \frac{J}{q} \,,
	\label{eqn:resinfty}
\end{equation}
we find that there are slivers of $\PT$ symmetric regions centered at modulation frequencies $\omega/J=2/n$ with odd $n$, deep in the otherwise $\PT$-symmetry broken landscape. In fact, motivated by this, we see that the substitution of Eq.~(\ref{eqn:resinfty}) into Eq.~(\ref{eqn:pt-h-cos}) results in
\begin{equation}
	\cos(2\varepsilon_F\tau) = \cos(J\tau)\, e^{-q\tau}
\end{equation}
exactly. The quantity on the right-hand side is always less than one; thus, the Floquet eigenvalues corresponding $\varepsilon_F$ are real along the line described by Eq.~(\ref{eqn:resinfty}) for all $\gamma_0>J$.

In Fig.~\ref{fig:ptphase-detail}(b), we show the $\PT$ phase diagram for this situation in the plane of $\omega$ and $\gamma_0$. We have drawn dashed white lines which correspond to Eq.~(\ref{eqn:ellipse}) at $\gamma_0/J \leq 1$, and the dashed lines which correspond to the solutions for $\gamma_0/J \gg 1$, given by Eq.~(\ref{eqn:resinfty}) are in red.

\section{Conclusion}
\label{sec:conclusion}

In conclusion we have presented a model which allows us to connect several active and passive time-periodic $\PT$-symmetric systems continuously to the static $\PT$ problem through a parameter $|\mu|\leq 1$. We have shown the progression of these models from the static case at $\mu = 1$, with a single $\PT$ symmetry breaking threshold to the case in which the system is driven from a $\PT$-symmetric Hamiltonian to a Hermitian one at $\mu=0$. 

This progression shows the introduction of many small regions of broken $\PT$ symmetry into the statically unbroken regime (with gain/loss $\gamma_0/J < 1$) which extend down to small neighborhoods around resonance frequencies $\omega/J = 2/n$. Likewise, regions of unbroken $\PT$ symmetry also extend upwards above $\gamma_0/J = 1$, so that at $\mu = 0$, they in fact reach arbitrarily far into the statically $\PT$-broken regime in small neighborhoods around $\omega/J = 2/n$ for odd $n$. For high driving frequencies, this system approaches that of a static $\PT$-symmetric system with an effective $\PT$-breaking threshold $\gamma_\infty/J = 2$, or twice that of the static threshold.

Advancing along this progression, from $\mu = 0$ to $\mu = -1$, we have seen that some of the regions of $\PT$ broken symmetry which had previously been introduced in the progression from $\mu = 1$ to $\mu = 0$ are then removed in this progression with $\mu < 0$. Specifically the $\PT$ broken regions corresponding to resonance frequencies $\omega/J = 2/n$ for even choices of $n$ begin to recede, and when $\mu = -1$, they have completely vanished. In this final case, in the high driving frequency limit, the system approaches a Hermitian one with a divergent $\gamma_\infty$. 
\begin{acknowledgments}
The authors acknowledge financial support from the following source: the Japan Society for the Promotion of Science (JSPS) Grants-in-Aid for Scientific Research (KAKENHI) JP19F19321 (A. H.). A. H. would also like to acknowledge support as an International Research Fellow of JSPS (Postdoctoral Fellowships for research in Japan (Standard)). This project began at the IIS-Chiba workshop NH2019TD and Y.J. is thankful to Prof. Naomichi Hatano for his hospitality.
\end{acknowledgments}

\bibliography{bib/journals, bib/citations}

\begin{thebibliography}{69}%
\makeatletter
\providecommand \@ifxundefined [1]{%
 \@ifx{#1\undefined}
}%
\providecommand \@ifnum [1]{%
 \ifnum #1\expandafter \@firstoftwo
 \else \expandafter \@secondoftwo
 \fi
}%
\providecommand \@ifx [1]{%
 \ifx #1\expandafter \@firstoftwo
 \else \expandafter \@secondoftwo
 \fi
}%
\providecommand \natexlab [1]{#1}%
\providecommand \enquote  [1]{``#1''}%
\providecommand \bibnamefont  [1]{#1}%
\providecommand \bibfnamefont [1]{#1}%
\providecommand \citenamefont [1]{#1}%
\providecommand \href@noop [0]{\@secondoftwo}%
\providecommand \href [0]{\begingroup \@sanitize@url \@href}%
\providecommand \@href[1]{\@@startlink{#1}\@@href}%
\providecommand \@@href[1]{\endgroup#1\@@endlink}%
\providecommand \@sanitize@url [0]{\catcode `\\12\catcode `\$12\catcode
  `\&12\catcode `\#12\catcode `\^12\catcode `\_12\catcode `\%12\relax}%
\providecommand \@@startlink[1]{}%
\providecommand \@@endlink[0]{}%
\providecommand \url  [0]{\begingroup\@sanitize@url \@url }%
\providecommand \@url [1]{\endgroup\@href {#1}{\urlprefix }}%
\providecommand \urlprefix  [0]{URL }%
\providecommand \Eprint [0]{\href }%
\providecommand \doibase [0]{http://dx.doi.org/}%
\providecommand \selectlanguage [0]{\@gobble}%
\providecommand \bibinfo  [0]{\@secondoftwo}%
\providecommand \bibfield  [0]{\@secondoftwo}%
\providecommand \translation [1]{[#1]}%
\providecommand \BibitemOpen [0]{}%
\providecommand \bibitemStop [0]{}%
\providecommand \bibitemNoStop [0]{.\EOS\space}%
\providecommand \EOS [0]{\spacefactor3000\relax}%
\providecommand \BibitemShut  [1]{\csname bibitem#1\endcsname}%
\let\auto@bib@innerbib\@empty
\bibitem [{\citenamefont {Bender}\ and\ \citenamefont
  {Boettcher}(1998)}]{Bender1998}%
  \BibitemOpen
  \bibfield  {author} {\bibinfo {author} {\bibfnamefont {C.~M.}\ \bibnamefont
  {Bender}}\ and\ \bibinfo {author} {\bibfnamefont {S.}~\bibnamefont
  {Boettcher}},\ }\href {\doibase 10.1103/physrevlett.80.5243} {\bibfield
  {journal} {\bibinfo  {journal} {Phys. Rev. Lett.}\ }\textbf {\bibinfo
  {volume} {80}},\ \bibinfo {pages} {5243} (\bibinfo {year}
  {1998})}\BibitemShut {NoStop}%
\bibitem [{\citenamefont {Bender}\ \emph {et~al.}(2002)\citenamefont {Bender},
  \citenamefont {Brody},\ and\ \citenamefont {Jones}}]{Bender2002}%
  \BibitemOpen
  \bibfield  {author} {\bibinfo {author} {\bibfnamefont {C.~M.}\ \bibnamefont
  {Bender}}, \bibinfo {author} {\bibfnamefont {D.~C.}\ \bibnamefont {Brody}}, \
  and\ \bibinfo {author} {\bibfnamefont {H.~F.}\ \bibnamefont {Jones}},\ }\href
  {\doibase 10.1103/physrevlett.89.270401} {\bibfield  {journal} {\bibinfo
  {journal} {Phys. Rev. Lett.}\ }\textbf {\bibinfo {volume} {89}},\ \bibinfo
  {pages} {270401} (\bibinfo {year} {2002})}\BibitemShut {NoStop}%
\bibitem [{\citenamefont {Bender}(2007)}]{Bender2007}%
  \BibitemOpen
  \bibfield  {author} {\bibinfo {author} {\bibfnamefont {C.~M.}\ \bibnamefont
  {Bender}},\ }\href {\doibase 10.1088/0034-4885/70/6/r03} {\bibfield
  {journal} {\bibinfo  {journal} {Rep. Prog. Phys.}\ }\textbf {\bibinfo
  {volume} {70}},\ \bibinfo {pages} {947} (\bibinfo {year} {2007})}\BibitemShut
  {NoStop}%
\bibitem [{\citenamefont {Heiss}(2004{\natexlab{a}})}]{Heiss2004a}%
  \BibitemOpen
  \bibfield  {author} {\bibinfo {author} {\bibfnamefont {W.~D.}\ \bibnamefont
  {Heiss}},\ }\href {\doibase 10.1088/0305-4470/37/6/034} {\bibfield  {journal}
  {\bibinfo  {journal} {J. Phys. A}\ }\textbf {\bibinfo {volume} {37}},\
  \bibinfo {pages} {2455} (\bibinfo {year} {2004}{\natexlab{a}})}\BibitemShut
  {NoStop}%
\bibitem [{\citenamefont {Heiss}(2004{\natexlab{b}})}]{Heiss2004b}%
  \BibitemOpen
  \bibfield  {author} {\bibinfo {author} {\bibfnamefont {W.~D.}\ \bibnamefont
  {Heiss}},\ }\href {\doibase 10.1023/B:CJOP.0000044009.17264.dc} {\bibfield
  {journal} {\bibinfo  {journal} {Czech. J. Phys.}\ }\textbf {\bibinfo {volume}
  {54}},\ \bibinfo {pages} {1091} (\bibinfo {year}
  {2004}{\natexlab{b}})}\BibitemShut {NoStop}%
\bibitem [{\citenamefont {Berry}(2004)}]{Berry2004}%
  \BibitemOpen
  \bibfield  {author} {\bibinfo {author} {\bibfnamefont {M.~V.}\ \bibnamefont
  {Berry}},\ }\href {\doibase 10.1023/B:CJOP.0000044002.05657.04} {\bibfield
  {journal} {\bibinfo  {journal} {Czech. J. Phys.}\ }\textbf {\bibinfo {volume}
  {54}},\ \bibinfo {pages} {1039} (\bibinfo {year} {2004})}\BibitemShut
  {NoStop}%
\bibitem [{\citenamefont {Hassan}\ \emph {et~al.}(2017)\citenamefont {Hassan},
  \citenamefont {Zhen}, \citenamefont {Solja\v{c}i\'c}, \citenamefont
  {Khajavikhan},\ and\ \citenamefont {Christodoulides}}]{Hassan2017}%
  \BibitemOpen
  \bibfield  {author} {\bibinfo {author} {\bibfnamefont {A.~U.}\ \bibnamefont
  {Hassan}}, \bibinfo {author} {\bibfnamefont {B.}~\bibnamefont {Zhen}},
  \bibinfo {author} {\bibfnamefont {M.}~\bibnamefont {Solja\v{c}i\'c}},
  \bibinfo {author} {\bibfnamefont {M.}~\bibnamefont {Khajavikhan}}, \ and\
  \bibinfo {author} {\bibfnamefont {D.~N.}\ \bibnamefont {Christodoulides}},\
  }\href {\doibase 10.1103/physrevlett.118.093002} {\bibfield  {journal}
  {\bibinfo  {journal} {Phys. Rev. Lett.}\ }\textbf {\bibinfo {volume} {118}},\
  \bibinfo {pages} {093002} (\bibinfo {year} {2017})}\BibitemShut {NoStop}%
\bibitem [{\citenamefont {Mostafazadeh}(2002)}]{Mostafazadeh2002}%
  \BibitemOpen
  \bibfield  {author} {\bibinfo {author} {\bibfnamefont {A.}~\bibnamefont
  {Mostafazadeh}},\ }\href {\doibase 10.1063/1.1418246} {\bibfield  {journal}
  {\bibinfo  {journal} {J. Math. Phys.}\ }\textbf {\bibinfo {volume} {43}},\
  \bibinfo {pages} {205} (\bibinfo {year} {2002})}\BibitemShut {NoStop}%
\bibitem [{\citenamefont {El-Ganainy}\ \emph {et~al.}(2007)\citenamefont
  {El-Ganainy}, \citenamefont {Makris}, \citenamefont {Christodoulides},\ and\
  \citenamefont {Musslimani}}]{ElGanainy2007}%
  \BibitemOpen
  \bibfield  {author} {\bibinfo {author} {\bibfnamefont {R.}~\bibnamefont
  {El-Ganainy}}, \bibinfo {author} {\bibfnamefont {K.~G.}\ \bibnamefont
  {Makris}}, \bibinfo {author} {\bibfnamefont {D.~N.}\ \bibnamefont
  {Christodoulides}}, \ and\ \bibinfo {author} {\bibfnamefont {Z.~H.}\
  \bibnamefont {Musslimani}},\ }\href {\doibase 10.1364/ol.32.002632}
  {\bibfield  {journal} {\bibinfo  {journal} {Opt. Lett.}\ }\textbf {\bibinfo
  {volume} {32}},\ \bibinfo {pages} {2632} (\bibinfo {year}
  {2007})}\BibitemShut {NoStop}%
\bibitem [{\citenamefont {Makris}\ \emph {et~al.}(2008)\citenamefont {Makris},
  \citenamefont {El-Ganainy}, \citenamefont {Christodoulides},\ and\
  \citenamefont {Musslimani}}]{Makris2008}%
  \BibitemOpen
  \bibfield  {author} {\bibinfo {author} {\bibfnamefont {K.~G.}\ \bibnamefont
  {Makris}}, \bibinfo {author} {\bibfnamefont {R.}~\bibnamefont {El-Ganainy}},
  \bibinfo {author} {\bibfnamefont {D.~N.}\ \bibnamefont {Christodoulides}}, \
  and\ \bibinfo {author} {\bibfnamefont {Z.~H.}\ \bibnamefont {Musslimani}},\
  }\href {\doibase 10.1103/physrevlett.100.103904} {\bibfield  {journal}
  {\bibinfo  {journal} {Phys. Rev. Lett.}\ }\textbf {\bibinfo {volume} {100}},\
  \bibinfo {pages} {103904} (\bibinfo {year} {2008})}\BibitemShut {NoStop}%
\bibitem [{\citenamefont {Guo}\ \emph {et~al.}(2009)\citenamefont {Guo},
  \citenamefont {Salamo}, \citenamefont {Duchesne}, \citenamefont {Morandotti},
  \citenamefont {Volatier-Ravat}, \citenamefont {Aimez}, \citenamefont
  {Siviloglou},\ and\ \citenamefont {Christodoulides}}]{Guo2009}%
  \BibitemOpen
  \bibfield  {author} {\bibinfo {author} {\bibfnamefont {A.}~\bibnamefont
  {Guo}}, \bibinfo {author} {\bibfnamefont {G.~J.}\ \bibnamefont {Salamo}},
  \bibinfo {author} {\bibfnamefont {D.}~\bibnamefont {Duchesne}}, \bibinfo
  {author} {\bibfnamefont {R.}~\bibnamefont {Morandotti}}, \bibinfo {author}
  {\bibfnamefont {M.}~\bibnamefont {Volatier-Ravat}}, \bibinfo {author}
  {\bibfnamefont {V.}~\bibnamefont {Aimez}}, \bibinfo {author} {\bibfnamefont
  {G.~A.}\ \bibnamefont {Siviloglou}}, \ and\ \bibinfo {author} {\bibfnamefont
  {D.~N.}\ \bibnamefont {Christodoulides}},\ }\href {\doibase
  10.1103/physrevlett.103.093902} {\bibfield  {journal} {\bibinfo  {journal}
  {Phys. Rev. Lett.}\ }\textbf {\bibinfo {volume} {103}},\ \bibinfo {pages}
  {093902} (\bibinfo {year} {2009})}\BibitemShut {NoStop}%
\bibitem [{\citenamefont {R\"uter}\ \emph {et~al.}(2010)\citenamefont
  {R\"uter}, \citenamefont {Makris}, \citenamefont {El-Ganainy}, \citenamefont
  {Christodoulides}, \citenamefont {Segev},\ and\ \citenamefont
  {Kip}}]{Ruter2010}%
  \BibitemOpen
  \bibfield  {author} {\bibinfo {author} {\bibfnamefont {C.~E.}\ \bibnamefont
  {R\"uter}}, \bibinfo {author} {\bibfnamefont {K.~G.}\ \bibnamefont {Makris}},
  \bibinfo {author} {\bibfnamefont {R.}~\bibnamefont {El-Ganainy}}, \bibinfo
  {author} {\bibfnamefont {D.~N.}\ \bibnamefont {Christodoulides}}, \bibinfo
  {author} {\bibfnamefont {M.}~\bibnamefont {Segev}}, \ and\ \bibinfo {author}
  {\bibfnamefont {D.}~\bibnamefont {Kip}},\ }\href {\doibase 10.1038/nphys1515}
  {\bibfield  {journal} {\bibinfo  {journal} {Nat. Phys.}\ }\textbf {\bibinfo
  {volume} {6}},\ \bibinfo {pages} {192} (\bibinfo {year} {2010})}\BibitemShut
  {NoStop}%
\bibitem [{\citenamefont {Szameit}\ \emph {et~al.}(2011)\citenamefont
  {Szameit}, \citenamefont {Rechtsman}, \citenamefont {Bahat-Treidel},\ and\
  \citenamefont {Segev}}]{Szameit2011}%
  \BibitemOpen
  \bibfield  {author} {\bibinfo {author} {\bibfnamefont {A.}~\bibnamefont
  {Szameit}}, \bibinfo {author} {\bibfnamefont {M.~C.}\ \bibnamefont
  {Rechtsman}}, \bibinfo {author} {\bibfnamefont {O.}~\bibnamefont
  {Bahat-Treidel}}, \ and\ \bibinfo {author} {\bibfnamefont {M.}~\bibnamefont
  {Segev}},\ }\href {\doibase 10.1103/physreva.84.021806} {\bibfield  {journal}
  {\bibinfo  {journal} {Phys. Rev. A}\ }\textbf {\bibinfo {volume} {84}},\
  \bibinfo {pages} {021806} (\bibinfo {year} {2011})}\BibitemShut {NoStop}%
\bibitem [{\citenamefont {Chang}\ \emph {et~al.}(2014)\citenamefont {Chang},
  \citenamefont {Jiang}, \citenamefont {Hua}, \citenamefont {Yang},
  \citenamefont {Wen}, \citenamefont {Jiang}, \citenamefont {Li}, \citenamefont
  {Wang},\ and\ \citenamefont {Xiao}}]{Chang2014}%
  \BibitemOpen
  \bibfield  {author} {\bibinfo {author} {\bibfnamefont {L.}~\bibnamefont
  {Chang}}, \bibinfo {author} {\bibfnamefont {X.}~\bibnamefont {Jiang}},
  \bibinfo {author} {\bibfnamefont {S.}~\bibnamefont {Hua}}, \bibinfo {author}
  {\bibfnamefont {C.}~\bibnamefont {Yang}}, \bibinfo {author} {\bibfnamefont
  {J.}~\bibnamefont {Wen}}, \bibinfo {author} {\bibfnamefont {L.}~\bibnamefont
  {Jiang}}, \bibinfo {author} {\bibfnamefont {G.}~\bibnamefont {Li}}, \bibinfo
  {author} {\bibfnamefont {G.}~\bibnamefont {Wang}}, \ and\ \bibinfo {author}
  {\bibfnamefont {M.}~\bibnamefont {Xiao}},\ }\href {\doibase
  10.1038/nphoton.2014.133} {\bibfield  {journal} {\bibinfo  {journal} {Nat.
  Photon.}\ }\textbf {\bibinfo {volume} {8}},\ \bibinfo {pages} {524} (\bibinfo
  {year} {2014})}\BibitemShut {NoStop}%
\bibitem [{\citenamefont {Hodaei}\ \emph {et~al.}(2014)\citenamefont {Hodaei},
  \citenamefont {Miri}, \citenamefont {Heinrich}, \citenamefont
  {Christodoulides},\ and\ \citenamefont {Khajavikhan}}]{Hodaei2014}%
  \BibitemOpen
  \bibfield  {author} {\bibinfo {author} {\bibfnamefont {H.}~\bibnamefont
  {Hodaei}}, \bibinfo {author} {\bibfnamefont {M.-A.}\ \bibnamefont {Miri}},
  \bibinfo {author} {\bibfnamefont {M.}~\bibnamefont {Heinrich}}, \bibinfo
  {author} {\bibfnamefont {D.~N.}\ \bibnamefont {Christodoulides}}, \ and\
  \bibinfo {author} {\bibfnamefont {M.}~\bibnamefont {Khajavikhan}},\ }\href
  {\doibase 10.1126/science.1258480} {\bibfield  {journal} {\bibinfo  {journal}
  {Science}\ }\textbf {\bibinfo {volume} {346}},\ \bibinfo {pages} {975}
  (\bibinfo {year} {2014})}\BibitemShut {NoStop}%
\bibitem [{\citenamefont {Peng}\ \emph {et~al.}(2014)\citenamefont {Peng},
  \citenamefont {\c{S}. K.~\"Ozdemir}, \citenamefont {Lei}, \citenamefont
  {Monifi}, \citenamefont {Gianfreda}, \citenamefont {Long}, \citenamefont
  {Fan}, \citenamefont {Nori}, \citenamefont {Bender},\ and\ \citenamefont
  {Yang}}]{Peng2014}%
  \BibitemOpen
  \bibfield  {author} {\bibinfo {author} {\bibfnamefont {B.}~\bibnamefont
  {Peng}}, \bibinfo {author} {\bibnamefont {\c{S}. K.~\"Ozdemir}}, \bibinfo
  {author} {\bibfnamefont {F.}~\bibnamefont {Lei}}, \bibinfo {author}
  {\bibfnamefont {F.}~\bibnamefont {Monifi}}, \bibinfo {author} {\bibfnamefont
  {M.}~\bibnamefont {Gianfreda}}, \bibinfo {author} {\bibfnamefont {G.~L.}\
  \bibnamefont {Long}}, \bibinfo {author} {\bibfnamefont {S.}~\bibnamefont
  {Fan}}, \bibinfo {author} {\bibfnamefont {F.}~\bibnamefont {Nori}}, \bibinfo
  {author} {\bibfnamefont {C.~M.}\ \bibnamefont {Bender}}, \ and\ \bibinfo
  {author} {\bibfnamefont {L.}~\bibnamefont {Yang}},\ }\href {\doibase
  10.1038/nphys2927} {\bibfield  {journal} {\bibinfo  {journal} {Nat. Phys.}\
  }\textbf {\bibinfo {volume} {10}},\ \bibinfo {pages} {394} (\bibinfo {year}
  {2014})}\BibitemShut {NoStop}%
\bibitem [{\citenamefont {Schindler}\ \emph {et~al.}(2011)\citenamefont
  {Schindler}, \citenamefont {Li}, \citenamefont {Zheng}, \citenamefont
  {Ellis},\ and\ \citenamefont {Kottos}}]{Schindler2011}%
  \BibitemOpen
  \bibfield  {author} {\bibinfo {author} {\bibfnamefont {J.}~\bibnamefont
  {Schindler}}, \bibinfo {author} {\bibfnamefont {A.}~\bibnamefont {Li}},
  \bibinfo {author} {\bibfnamefont {M.~C.}\ \bibnamefont {Zheng}}, \bibinfo
  {author} {\bibfnamefont {F.~M.}\ \bibnamefont {Ellis}}, \ and\ \bibinfo
  {author} {\bibfnamefont {T.}~\bibnamefont {Kottos}},\ }\href {\doibase
  10.1103/physreva.84.040101} {\bibfield  {journal} {\bibinfo  {journal} {Phys.
  Rev. A}\ }\textbf {\bibinfo {volume} {84}},\ \bibinfo {pages} {040101}
  (\bibinfo {year} {2011})}\BibitemShut {NoStop}%
\bibitem [{\citenamefont {Bender}\ \emph {et~al.}(2013)\citenamefont {Bender},
  \citenamefont {Berntson}, \citenamefont {Parker},\ and\ \citenamefont
  {Samuel}}]{Bender2013}%
  \BibitemOpen
  \bibfield  {author} {\bibinfo {author} {\bibfnamefont {C.~M.}\ \bibnamefont
  {Bender}}, \bibinfo {author} {\bibfnamefont {B.~K.}\ \bibnamefont
  {Berntson}}, \bibinfo {author} {\bibfnamefont {D.}~\bibnamefont {Parker}}, \
  and\ \bibinfo {author} {\bibfnamefont {E.}~\bibnamefont {Samuel}},\ }\href
  {\doibase 10.1119/1.4789549} {\bibfield  {journal} {\bibinfo  {journal} {Am.
  J. Phys.}\ }\textbf {\bibinfo {volume} {81}},\ \bibinfo {pages} {173}
  (\bibinfo {year} {2013})}\BibitemShut {NoStop}%
\bibitem [{\citenamefont {Fleury}\ \emph {et~al.}(2015)\citenamefont {Fleury},
  \citenamefont {Sounas},\ and\ \citenamefont {Al\`u}}]{Fleury2015}%
  \BibitemOpen
  \bibfield  {author} {\bibinfo {author} {\bibfnamefont {R.}~\bibnamefont
  {Fleury}}, \bibinfo {author} {\bibfnamefont {D.}~\bibnamefont {Sounas}}, \
  and\ \bibinfo {author} {\bibfnamefont {A.}~\bibnamefont {Al\`u}},\ }\href
  {\doibase 10.1038/ncomms6905} {\bibfield  {journal} {\bibinfo  {journal}
  {Nat. Commun.}\ }\textbf {\bibinfo {volume} {6}},\ \bibinfo {pages} {5905}
  (\bibinfo {year} {2015})}\BibitemShut {NoStop}%
\bibitem [{\citenamefont {Zhang}\ \emph {et~al.}(2016)\citenamefont {Zhang},
  \citenamefont {Zhang}, \citenamefont {Sheng}, \citenamefont {Yang},
  \citenamefont {Miri}, \citenamefont {Christodoulides}, \citenamefont {He},
  \citenamefont {Zhang},\ and\ \citenamefont {Xiao}}]{Zhang2016}%
  \BibitemOpen
  \bibfield  {author} {\bibinfo {author} {\bibfnamefont {Z.}~\bibnamefont
  {Zhang}}, \bibinfo {author} {\bibfnamefont {Y.}~\bibnamefont {Zhang}},
  \bibinfo {author} {\bibfnamefont {J.}~\bibnamefont {Sheng}}, \bibinfo
  {author} {\bibfnamefont {L.}~\bibnamefont {Yang}}, \bibinfo {author}
  {\bibfnamefont {M.-A.}\ \bibnamefont {Miri}}, \bibinfo {author}
  {\bibfnamefont {D.~N.}\ \bibnamefont {Christodoulides}}, \bibinfo {author}
  {\bibfnamefont {B.}~\bibnamefont {He}}, \bibinfo {author} {\bibfnamefont
  {Y.}~\bibnamefont {Zhang}}, \ and\ \bibinfo {author} {\bibfnamefont
  {M.}~\bibnamefont {Xiao}},\ }\href {\doibase 10.1103/physrevlett.117.123601}
  {\bibfield  {journal} {\bibinfo  {journal} {Phys. Rev. Lett.}\ }\textbf
  {\bibinfo {volume} {117}},\ \bibinfo {pages} {123601} (\bibinfo {year}
  {2016})}\BibitemShut {NoStop}%
\bibitem [{\citenamefont {Peng}\ \emph {et~al.}(2016)\citenamefont {Peng},
  \citenamefont {Cao}, \citenamefont {Shen}, \citenamefont {Qu}, \citenamefont
  {Wen}, \citenamefont {Jiang},\ and\ \citenamefont {Xiao}}]{Peng2016}%
  \BibitemOpen
  \bibfield  {author} {\bibinfo {author} {\bibfnamefont {P.}~\bibnamefont
  {Peng}}, \bibinfo {author} {\bibfnamefont {W.}~\bibnamefont {Cao}}, \bibinfo
  {author} {\bibfnamefont {C.}~\bibnamefont {Shen}}, \bibinfo {author}
  {\bibfnamefont {W.}~\bibnamefont {Qu}}, \bibinfo {author} {\bibfnamefont
  {J.}~\bibnamefont {Wen}}, \bibinfo {author} {\bibfnamefont {L.}~\bibnamefont
  {Jiang}}, \ and\ \bibinfo {author} {\bibfnamefont {Y.}~\bibnamefont {Xiao}},\
  }\href {\doibase 10.1038/nphys3842} {\bibfield  {journal} {\bibinfo
  {journal} {Nat. Phys.}\ }\textbf {\bibinfo {volume} {12}},\ \bibinfo {pages}
  {1139} (\bibinfo {year} {2016})}\BibitemShut {NoStop}%
\bibitem [{\citenamefont {Xiao}\ \emph {et~al.}(2017)\citenamefont {Xiao},
  \citenamefont {Zhan}, \citenamefont {Bian}, \citenamefont {Wang},
  \citenamefont {Zhang}, \citenamefont {Wang}, \citenamefont {Li},
  \citenamefont {Mochizuki}, \citenamefont {Kim}, \citenamefont {Kawakami},
  \citenamefont {Yi}, \citenamefont {Obuse}, \citenamefont {Sanders},\ and\
  \citenamefont {Xue}}]{Xiao2017}%
  \BibitemOpen
  \bibfield  {author} {\bibinfo {author} {\bibfnamefont {L.}~\bibnamefont
  {Xiao}}, \bibinfo {author} {\bibfnamefont {X.}~\bibnamefont {Zhan}}, \bibinfo
  {author} {\bibfnamefont {Z.~H.}\ \bibnamefont {Bian}}, \bibinfo {author}
  {\bibfnamefont {K.~K.}\ \bibnamefont {Wang}}, \bibinfo {author}
  {\bibfnamefont {X.}~\bibnamefont {Zhang}}, \bibinfo {author} {\bibfnamefont
  {X.~P.}\ \bibnamefont {Wang}}, \bibinfo {author} {\bibfnamefont
  {J.}~\bibnamefont {Li}}, \bibinfo {author} {\bibfnamefont {K.}~\bibnamefont
  {Mochizuki}}, \bibinfo {author} {\bibfnamefont {D.}~\bibnamefont {Kim}},
  \bibinfo {author} {\bibfnamefont {N.}~\bibnamefont {Kawakami}}, \bibinfo
  {author} {\bibfnamefont {W.}~\bibnamefont {Yi}}, \bibinfo {author}
  {\bibfnamefont {H.}~\bibnamefont {Obuse}}, \bibinfo {author} {\bibfnamefont
  {B.~C.}\ \bibnamefont {Sanders}}, \ and\ \bibinfo {author} {\bibfnamefont
  {P.}~\bibnamefont {Xue}},\ }\href {\doibase 10.1038/nphys4204} {\bibfield
  {journal} {\bibinfo  {journal} {Nat. Phys.}\ }\textbf {\bibinfo {volume}
  {13}},\ \bibinfo {pages} {1117} (\bibinfo {year} {2017})}\BibitemShut
  {NoStop}%
\bibitem [{\citenamefont {Tang}\ \emph {et~al.}(2016)\citenamefont {Tang},
  \citenamefont {Wang}, \citenamefont {Yu}, \citenamefont {He}, \citenamefont
  {Xu}, \citenamefont {Liu}, \citenamefont {Chen}, \citenamefont {Sun},
  \citenamefont {Sun}, \citenamefont {Han}, \citenamefont {Li},\ and\
  \citenamefont {Guo}}]{Tang2016}%
  \BibitemOpen
  \bibfield  {author} {\bibinfo {author} {\bibfnamefont {J.-S.}\ \bibnamefont
  {Tang}}, \bibinfo {author} {\bibfnamefont {Y.-T.}\ \bibnamefont {Wang}},
  \bibinfo {author} {\bibfnamefont {S.}~\bibnamefont {Yu}}, \bibinfo {author}
  {\bibfnamefont {D.-Y.}\ \bibnamefont {He}}, \bibinfo {author} {\bibfnamefont
  {J.-S.}\ \bibnamefont {Xu}}, \bibinfo {author} {\bibfnamefont {B.-H.}\
  \bibnamefont {Liu}}, \bibinfo {author} {\bibfnamefont {G.}~\bibnamefont
  {Chen}}, \bibinfo {author} {\bibfnamefont {Y.-N.}\ \bibnamefont {Sun}},
  \bibinfo {author} {\bibfnamefont {K.}~\bibnamefont {Sun}}, \bibinfo {author}
  {\bibfnamefont {Y.-J.}\ \bibnamefont {Han}}, \bibinfo {author} {\bibfnamefont
  {C.-F.}\ \bibnamefont {Li}}, \ and\ \bibinfo {author} {\bibfnamefont {G.-C.}\
  \bibnamefont {Guo}},\ }\href {\doibase 10.1038/nphoton.2016.144} {\bibfield
  {journal} {\bibinfo  {journal} {Nat. Photon.}\ }\textbf {\bibinfo {volume}
  {10}},\ \bibinfo {pages} {642} (\bibinfo {year} {2016})}\BibitemShut
  {NoStop}%
\bibitem [{\citenamefont {Klauck}\ \emph {et~al.}(2019)\citenamefont {Klauck},
  \citenamefont {Teuber}, \citenamefont {Ornigotti}, \citenamefont {Heinrich},
  \citenamefont {Scheel},\ and\ \citenamefont {Szameit}}]{Klauck2019}%
  \BibitemOpen
  \bibfield  {author} {\bibinfo {author} {\bibfnamefont {F.}~\bibnamefont
  {Klauck}}, \bibinfo {author} {\bibfnamefont {L.}~\bibnamefont {Teuber}},
  \bibinfo {author} {\bibfnamefont {M.}~\bibnamefont {Ornigotti}}, \bibinfo
  {author} {\bibfnamefont {M.}~\bibnamefont {Heinrich}}, \bibinfo {author}
  {\bibfnamefont {S.}~\bibnamefont {Scheel}}, \ and\ \bibinfo {author}
  {\bibfnamefont {A.}~\bibnamefont {Szameit}},\ }\href {\doibase
  10.1038/s41566-019-0517-0} {\bibfield  {journal} {\bibinfo  {journal} {Nat.
  Photon.}\ }\textbf {\bibinfo {volume} {13}},\ \bibinfo {pages} {883}
  (\bibinfo {year} {2019})}\BibitemShut {NoStop}%
\bibitem [{\citenamefont {Xiao}\ \emph {et~al.}(2019)\citenamefont {Xiao},
  \citenamefont {Wang}, \citenamefont {Zhan}, \citenamefont {Bian},
  \citenamefont {Kawabata}, \citenamefont {Ueda}, \citenamefont {Yi},\ and\
  \citenamefont {Xue}}]{Xiao2019}%
  \BibitemOpen
  \bibfield  {author} {\bibinfo {author} {\bibfnamefont {L.}~\bibnamefont
  {Xiao}}, \bibinfo {author} {\bibfnamefont {K.}~\bibnamefont {Wang}}, \bibinfo
  {author} {\bibfnamefont {X.}~\bibnamefont {Zhan}}, \bibinfo {author}
  {\bibfnamefont {Z.}~\bibnamefont {Bian}}, \bibinfo {author} {\bibfnamefont
  {K.}~\bibnamefont {Kawabata}}, \bibinfo {author} {\bibfnamefont
  {M.}~\bibnamefont {Ueda}}, \bibinfo {author} {\bibfnamefont {W.}~\bibnamefont
  {Yi}}, \ and\ \bibinfo {author} {\bibfnamefont {P.}~\bibnamefont {Xue}},\
  }\href {\doibase 10.1103/PhysRevLett.123.230401} {\bibfield  {journal}
  {\bibinfo  {journal} {Phys. Rev. Lett.}\ }\textbf {\bibinfo {volume} {123}},\
  \bibinfo {pages} {230401} (\bibinfo {year} {2019})}\BibitemShut {NoStop}%
\bibitem [{\citenamefont {Bian}\ \emph {et~al.}(2020)\citenamefont {Bian},
  \citenamefont {Xiao}, \citenamefont {Wang}, \citenamefont {Zhan},
  \citenamefont {{Assogba Onanga}}, \citenamefont {Ruzicka}, \citenamefont
  {Yi}, \citenamefont {Joglekar},\ and\ \citenamefont {Xue}}]{Bian2020}%
  \BibitemOpen
  \bibfield  {author} {\bibinfo {author} {\bibfnamefont {Z.}~\bibnamefont
  {Bian}}, \bibinfo {author} {\bibfnamefont {L.}~\bibnamefont {Xiao}}, \bibinfo
  {author} {\bibfnamefont {K.}~\bibnamefont {Wang}}, \bibinfo {author}
  {\bibfnamefont {X.}~\bibnamefont {Zhan}}, \bibinfo {author} {\bibfnamefont
  {F.}~\bibnamefont {{Assogba Onanga}}}, \bibinfo {author} {\bibfnamefont
  {F.}~\bibnamefont {Ruzicka}}, \bibinfo {author} {\bibfnamefont
  {W.}~\bibnamefont {Yi}}, \bibinfo {author} {\bibfnamefont {Y.~N.}\
  \bibnamefont {Joglekar}}, \ and\ \bibinfo {author} {\bibfnamefont
  {P.}~\bibnamefont {Xue}},\ }\href {\doibase 10.1103/PhysRevResearch.2.022039}
  {\bibfield  {journal} {\bibinfo  {journal} {Phys. Rev. Res.}\ }\textbf
  {\bibinfo {volume} {2}},\ \bibinfo {pages} {022039(R)} (\bibinfo {year}
  {2020})}\BibitemShut {NoStop}%
\bibitem [{\citenamefont {Naghiloo}\ \emph {et~al.}(2019)\citenamefont
  {Naghiloo}, \citenamefont {Abbasi}, \citenamefont {Joglekar},\ and\
  \citenamefont {Murch}}]{Naghiloo2019}%
  \BibitemOpen
  \bibfield  {author} {\bibinfo {author} {\bibfnamefont {M.}~\bibnamefont
  {Naghiloo}}, \bibinfo {author} {\bibfnamefont {M.}~\bibnamefont {Abbasi}},
  \bibinfo {author} {\bibfnamefont {Y.~N.}\ \bibnamefont {Joglekar}}, \ and\
  \bibinfo {author} {\bibfnamefont {K.~W.}\ \bibnamefont {Murch}},\ }\href
  {\doibase 10.1038/s41567-019-0652-z} {\bibfield  {journal} {\bibinfo
  {journal} {Nat. Phys.}\ }\textbf {\bibinfo {volume} {15}},\ \bibinfo {pages}
  {1232} (\bibinfo {year} {2019})}\BibitemShut {NoStop}%
\bibitem [{\citenamefont {Wu}\ \emph {et~al.}(2019)\citenamefont {Wu},
  \citenamefont {Liu}, \citenamefont {Geng}, \citenamefont {Song},
  \citenamefont {Ye}, \citenamefont {Duan}, \citenamefont {Rong},\ and\
  \citenamefont {Du}}]{Wu2019}%
  \BibitemOpen
  \bibfield  {author} {\bibinfo {author} {\bibfnamefont {Y.}~\bibnamefont
  {Wu}}, \bibinfo {author} {\bibfnamefont {W.}~\bibnamefont {Liu}}, \bibinfo
  {author} {\bibfnamefont {J.}~\bibnamefont {Geng}}, \bibinfo {author}
  {\bibfnamefont {X.}~\bibnamefont {Song}}, \bibinfo {author} {\bibfnamefont
  {X.}~\bibnamefont {Ye}}, \bibinfo {author} {\bibfnamefont {C.-K.}\
  \bibnamefont {Duan}}, \bibinfo {author} {\bibfnamefont {X.}~\bibnamefont
  {Rong}}, \ and\ \bibinfo {author} {\bibfnamefont {J.}~\bibnamefont {Du}},\
  }\href {\doibase 10.1126/science.aaw8205} {\bibfield  {journal} {\bibinfo
  {journal} {Science}\ }\textbf {\bibinfo {volume} {364}},\ \bibinfo {pages}
  {878} (\bibinfo {year} {2019})}\BibitemShut {NoStop}%
\bibitem [{\citenamefont {Zheng}\ \emph {et~al.}(2013)\citenamefont {Zheng},
  \citenamefont {Hao},\ and\ \citenamefont {Long}}]{Zheng2013}%
  \BibitemOpen
  \bibfield  {author} {\bibinfo {author} {\bibfnamefont {C.}~\bibnamefont
  {Zheng}}, \bibinfo {author} {\bibfnamefont {L.}~\bibnamefont {Hao}}, \ and\
  \bibinfo {author} {\bibfnamefont {G.~L.}\ \bibnamefont {Long}},\ }\href
  {\doibase 10.1098/rsta.2012.0053} {\bibfield  {journal} {\bibinfo  {journal}
  {Phil. Trans. R. Soc. A}\ }\textbf {\bibinfo {volume} {371}},\ \bibinfo
  {pages} {20120053} (\bibinfo {year} {2013})}\BibitemShut {NoStop}%
\bibitem [{\citenamefont {Wen}\ \emph {et~al.}(2019)\citenamefont {Wen},
  \citenamefont {Zheng}, \citenamefont {Kong}, \citenamefont {Wei},
  \citenamefont {Xin},\ and\ \citenamefont {Long}}]{Wen2019}%
  \BibitemOpen
  \bibfield  {author} {\bibinfo {author} {\bibfnamefont {J.}~\bibnamefont
  {Wen}}, \bibinfo {author} {\bibfnamefont {C.}~\bibnamefont {Zheng}}, \bibinfo
  {author} {\bibfnamefont {X.}~\bibnamefont {Kong}}, \bibinfo {author}
  {\bibfnamefont {S.}~\bibnamefont {Wei}}, \bibinfo {author} {\bibfnamefont
  {T.}~\bibnamefont {Xin}}, \ and\ \bibinfo {author} {\bibfnamefont
  {G.}~\bibnamefont {Long}},\ }\href {\doibase 10.1103/physreva.99.062122}
  {\bibfield  {journal} {\bibinfo  {journal} {Phys. Rev. A}\ }\textbf {\bibinfo
  {volume} {99}},\ \bibinfo {pages} {062122} (\bibinfo {year}
  {2019})}\BibitemShut {NoStop}%
\bibitem [{\citenamefont {Joglekar}\ \emph {et~al.}(2013)\citenamefont
  {Joglekar}, \citenamefont {Thompson}, \citenamefont {Scott},\ and\
  \citenamefont {Vemuri}}]{Joglekar2013}%
  \BibitemOpen
  \bibfield  {author} {\bibinfo {author} {\bibfnamefont {Y.~N.}\ \bibnamefont
  {Joglekar}}, \bibinfo {author} {\bibfnamefont {C.}~\bibnamefont {Thompson}},
  \bibinfo {author} {\bibfnamefont {D.~D.}\ \bibnamefont {Scott}}, \ and\
  \bibinfo {author} {\bibfnamefont {G.}~\bibnamefont {Vemuri}},\ }\href
  {\doibase 10.1051/epjap/2013130240} {\bibfield  {journal} {\bibinfo
  {journal} {Eur. Phys. J. Appl. Phys.}\ }\textbf {\bibinfo {volume} {63}},\
  \bibinfo {pages} {30001} (\bibinfo {year} {2013})}\BibitemShut {NoStop}%
\bibitem [{\citenamefont {Feng}\ \emph {et~al.}(2017)\citenamefont {Feng},
  \citenamefont {El-Ganainy},\ and\ \citenamefont {Ge}}]{Feng2017}%
  \BibitemOpen
  \bibfield  {author} {\bibinfo {author} {\bibfnamefont {L.}~\bibnamefont
  {Feng}}, \bibinfo {author} {\bibfnamefont {R.}~\bibnamefont {El-Ganainy}}, \
  and\ \bibinfo {author} {\bibfnamefont {L.}~\bibnamefont {Ge}},\ }\href
  {\doibase 10.1038/s41566-017-0031-1} {\bibfield  {journal} {\bibinfo
  {journal} {Nat. Photon.}\ }\textbf {\bibinfo {volume} {11}},\ \bibinfo
  {pages} {752} (\bibinfo {year} {2017})}\BibitemShut {NoStop}%
\bibitem [{\citenamefont {El-Ganainy}\ \emph {et~al.}(2018)\citenamefont
  {El-Ganainy}, \citenamefont {Makris}, \citenamefont {Khajavikhan},
  \citenamefont {Musslimani}, \citenamefont {Rotter},\ and\ \citenamefont
  {Christodoulides}}]{ElGanainy2018}%
  \BibitemOpen
  \bibfield  {author} {\bibinfo {author} {\bibfnamefont {R.}~\bibnamefont
  {El-Ganainy}}, \bibinfo {author} {\bibfnamefont {K.~G.}\ \bibnamefont
  {Makris}}, \bibinfo {author} {\bibfnamefont {M.}~\bibnamefont {Khajavikhan}},
  \bibinfo {author} {\bibfnamefont {Z.~H.}\ \bibnamefont {Musslimani}},
  \bibinfo {author} {\bibfnamefont {S.}~\bibnamefont {Rotter}}, \ and\ \bibinfo
  {author} {\bibfnamefont {D.~N.}\ \bibnamefont {Christodoulides}},\ }\href
  {\doibase 10.1038/nphys4323} {\bibfield  {journal} {\bibinfo  {journal} {Nat.
  Phys.}\ }\textbf {\bibinfo {volume} {14}},\ \bibinfo {pages} {11} (\bibinfo
  {year} {2018})}\BibitemShut {NoStop}%
\bibitem [{\citenamefont {\c{S}. K.~\"Ozdemir}\ \emph
  {et~al.}(2019)\citenamefont {\c{S}. K.~\"Ozdemir}, \citenamefont {Rotter},
  \citenamefont {Nori},\ and\ \citenamefont {Yang}}]{Ozdemir2019}%
  \BibitemOpen
  \bibfield  {author} {\bibinfo {author} {\bibnamefont {\c{S}. K.~\"Ozdemir}},
  \bibinfo {author} {\bibfnamefont {S.}~\bibnamefont {Rotter}}, \bibinfo
  {author} {\bibfnamefont {F.}~\bibnamefont {Nori}}, \ and\ \bibinfo {author}
  {\bibfnamefont {L.}~\bibnamefont {Yang}},\ }\href {\doibase
  10.1038/s41563-019-0304-9} {\bibfield  {journal} {\bibinfo  {journal} {Nat.
  Mater.}\ }\textbf {\bibinfo {volume} {18}},\ \bibinfo {pages} {783} (\bibinfo
  {year} {2019})}\BibitemShut {NoStop}%
\bibitem [{\citenamefont {Floquet}(1883)}]{Floquet1883}%
  \BibitemOpen
  \bibfield  {author} {\bibinfo {author} {\bibfnamefont {G.}~\bibnamefont
  {Floquet}},\ }\href {\doibase 10.24033/asens.220} {\bibfield  {journal}
  {\bibinfo  {journal} {Ann. Sci. \'Ec. Norm. Sup\'er.}\ }\textbf {\bibinfo
  {volume} {12}},\ \bibinfo {pages} {47} (\bibinfo {year} {1883})}\BibitemShut
  {NoStop}%
\bibitem [{\citenamefont {Shirley}(1965)}]{Shirley1965}%
  \BibitemOpen
  \bibfield  {author} {\bibinfo {author} {\bibfnamefont {J.~H.}\ \bibnamefont
  {Shirley}},\ }\href {\doibase 10.1103/physrev.138.b979} {\bibfield  {journal}
  {\bibinfo  {journal} {Phys. Rev.}\ }\textbf {\bibinfo {volume} {138}},\
  \bibinfo {pages} {B979} (\bibinfo {year} {1965})}\BibitemShut {NoStop}%
\bibitem [{\citenamefont {Kitagawa}\ \emph {et~al.}(2010)\citenamefont
  {Kitagawa}, \citenamefont {Berg}, \citenamefont {Rudner},\ and\ \citenamefont
  {Demler}}]{Kitagawa2010}%
  \BibitemOpen
  \bibfield  {author} {\bibinfo {author} {\bibfnamefont {T.}~\bibnamefont
  {Kitagawa}}, \bibinfo {author} {\bibfnamefont {E.}~\bibnamefont {Berg}},
  \bibinfo {author} {\bibfnamefont {M.}~\bibnamefont {Rudner}}, \ and\ \bibinfo
  {author} {\bibfnamefont {E.}~\bibnamefont {Demler}},\ }\href {\doibase
  10.1103/physrevb.82.235114} {\bibfield  {journal} {\bibinfo  {journal} {Phys.
  Rev. B}\ }\textbf {\bibinfo {volume} {82}},\ \bibinfo {pages} {235114}
  (\bibinfo {year} {2010})}\BibitemShut {NoStop}%
\bibitem [{\citenamefont {{Dal Lago}}\ \emph {et~al.}(2015)\citenamefont {{Dal
  Lago}}, \citenamefont {Atala},\ and\ \citenamefont {{Foa
  Torres}}}]{DalLago2015}%
  \BibitemOpen
  \bibfield  {author} {\bibinfo {author} {\bibfnamefont {V.}~\bibnamefont {{Dal
  Lago}}}, \bibinfo {author} {\bibfnamefont {M.}~\bibnamefont {Atala}}, \ and\
  \bibinfo {author} {\bibfnamefont {L.~E.~F.}\ \bibnamefont {{Foa Torres}}},\
  }\href {\doibase 10.1103/physreva.92.023624} {\bibfield  {journal} {\bibinfo
  {journal} {Phys. Rev. A}\ }\textbf {\bibinfo {volume} {92}},\ \bibinfo
  {pages} {023624} (\bibinfo {year} {2015})}\BibitemShut {NoStop}%
\bibitem [{\citenamefont {Fruchart}(2016)}]{Fruchart2016}%
  \BibitemOpen
  \bibfield  {author} {\bibinfo {author} {\bibfnamefont {M.}~\bibnamefont
  {Fruchart}},\ }\href {\doibase 10.1103/physrevb.93.115429} {\bibfield
  {journal} {\bibinfo  {journal} {Phys. Rev. B}\ }\textbf {\bibinfo {volume}
  {93}},\ \bibinfo {pages} {115429} (\bibinfo {year} {2016})}\BibitemShut
  {NoStop}%
\bibitem [{\citenamefont {Rechtsman}\ \emph {et~al.}(2013)\citenamefont
  {Rechtsman}, \citenamefont {Zeuner}, \citenamefont {Plotnik}, \citenamefont
  {Lumer}, \citenamefont {Podolsky}, \citenamefont {Dreisow}, \citenamefont
  {Nolte}, \citenamefont {Segev},\ and\ \citenamefont
  {Szameit}}]{Rechtsman2013}%
  \BibitemOpen
  \bibfield  {author} {\bibinfo {author} {\bibfnamefont {M.~C.}\ \bibnamefont
  {Rechtsman}}, \bibinfo {author} {\bibfnamefont {J.~M.}\ \bibnamefont
  {Zeuner}}, \bibinfo {author} {\bibfnamefont {Y.}~\bibnamefont {Plotnik}},
  \bibinfo {author} {\bibfnamefont {Y.}~\bibnamefont {Lumer}}, \bibinfo
  {author} {\bibfnamefont {D.}~\bibnamefont {Podolsky}}, \bibinfo {author}
  {\bibfnamefont {F.}~\bibnamefont {Dreisow}}, \bibinfo {author} {\bibfnamefont
  {S.}~\bibnamefont {Nolte}}, \bibinfo {author} {\bibfnamefont
  {M.}~\bibnamefont {Segev}}, \ and\ \bibinfo {author} {\bibfnamefont
  {A.}~\bibnamefont {Szameit}},\ }\href {\doibase 10.1038/nature12066}
  {\bibfield  {journal} {\bibinfo  {journal} {Nature}\ }\textbf {\bibinfo
  {volume} {496}},\ \bibinfo {pages} {196} (\bibinfo {year}
  {2013})}\BibitemShut {NoStop}%
\bibitem [{\citenamefont {Wang}\ \emph {et~al.}(2013)\citenamefont {Wang},
  \citenamefont {Steinberg}, \citenamefont {Jarillo-Herrero},\ and\
  \citenamefont {Gedik}}]{Wang2013}%
  \BibitemOpen
  \bibfield  {author} {\bibinfo {author} {\bibfnamefont {Y.~H.}\ \bibnamefont
  {Wang}}, \bibinfo {author} {\bibfnamefont {H.}~\bibnamefont {Steinberg}},
  \bibinfo {author} {\bibfnamefont {P.}~\bibnamefont {Jarillo-Herrero}}, \ and\
  \bibinfo {author} {\bibfnamefont {N.}~\bibnamefont {Gedik}},\ }\href
  {\doibase 10.1126/science.1239834} {\bibfield  {journal} {\bibinfo  {journal}
  {Science}\ }\textbf {\bibinfo {volume} {342}},\ \bibinfo {pages} {453}
  (\bibinfo {year} {2013})}\BibitemShut {NoStop}%
\bibitem [{\citenamefont {Jotzu}\ \emph {et~al.}(2014)\citenamefont {Jotzu},
  \citenamefont {Messer}, \citenamefont {Desbuquois}, \citenamefont {Lebrat},
  \citenamefont {Uehlinger}, \citenamefont {Greif},\ and\ \citenamefont
  {Esslinger}}]{Jotzu2014}%
  \BibitemOpen
  \bibfield  {author} {\bibinfo {author} {\bibfnamefont {G.}~\bibnamefont
  {Jotzu}}, \bibinfo {author} {\bibfnamefont {M.}~\bibnamefont {Messer}},
  \bibinfo {author} {\bibfnamefont {R.}~\bibnamefont {Desbuquois}}, \bibinfo
  {author} {\bibfnamefont {M.}~\bibnamefont {Lebrat}}, \bibinfo {author}
  {\bibfnamefont {T.}~\bibnamefont {Uehlinger}}, \bibinfo {author}
  {\bibfnamefont {D.}~\bibnamefont {Greif}}, \ and\ \bibinfo {author}
  {\bibfnamefont {T.}~\bibnamefont {Esslinger}},\ }\href {\doibase
  10.1038/nature13915} {\bibfield  {journal} {\bibinfo  {journal} {Nature}\
  }\textbf {\bibinfo {volume} {515}},\ \bibinfo {pages} {237} (\bibinfo {year}
  {2014})}\BibitemShut {NoStop}%
\bibitem [{\citenamefont {Maczewsky}\ \emph {et~al.}(2017)\citenamefont
  {Maczewsky}, \citenamefont {Zeuner}, \citenamefont {Nolte},\ and\
  \citenamefont {Szameit}}]{Maczewsky2017}%
  \BibitemOpen
  \bibfield  {author} {\bibinfo {author} {\bibfnamefont {L.~J.}\ \bibnamefont
  {Maczewsky}}, \bibinfo {author} {\bibfnamefont {J.~M.}\ \bibnamefont
  {Zeuner}}, \bibinfo {author} {\bibfnamefont {S.}~\bibnamefont {Nolte}}, \
  and\ \bibinfo {author} {\bibfnamefont {A.}~\bibnamefont {Szameit}},\ }\href
  {\doibase 10.1038/ncomms13756} {\bibfield  {journal} {\bibinfo  {journal}
  {Nat. Commun.}\ }\textbf {\bibinfo {volume} {8}},\ \bibinfo {pages} {13756}
  (\bibinfo {year} {2017})}\BibitemShut {NoStop}%
\bibitem [{\citenamefont {Bordia}\ \emph {et~al.}(2017)\citenamefont {Bordia},
  \citenamefont {L\"uschen}, \citenamefont {Schneider}, \citenamefont {Knap},\
  and\ \citenamefont {Bloch}}]{Bordia2017}%
  \BibitemOpen
  \bibfield  {author} {\bibinfo {author} {\bibfnamefont {P.}~\bibnamefont
  {Bordia}}, \bibinfo {author} {\bibfnamefont {H.}~\bibnamefont {L\"uschen}},
  \bibinfo {author} {\bibfnamefont {U.}~\bibnamefont {Schneider}}, \bibinfo
  {author} {\bibfnamefont {M.}~\bibnamefont {Knap}}, \ and\ \bibinfo {author}
  {\bibfnamefont {I.}~\bibnamefont {Bloch}},\ }\href {\doibase
  10.1038/nphys4020} {\bibfield  {journal} {\bibinfo  {journal} {Nat. Phys.}\
  }\textbf {\bibinfo {volume} {13}},\ \bibinfo {pages} {460} (\bibinfo {year}
  {2017})}\BibitemShut {NoStop}%
\bibitem [{\citenamefont {Wintersperger}\ \emph {et~al.}(2020)\citenamefont
  {Wintersperger}, \citenamefont {Braun}, \citenamefont {\"Unal}, \citenamefont
  {Eckardt}, \citenamefont {{Di Liberto}}, \citenamefont {Goldman},
  \citenamefont {Bloch},\ and\ \citenamefont
  {Aidelsburger}}]{Wintersperger2020}%
  \BibitemOpen
  \bibfield  {author} {\bibinfo {author} {\bibfnamefont {K.}~\bibnamefont
  {Wintersperger}}, \bibinfo {author} {\bibfnamefont {C.}~\bibnamefont
  {Braun}}, \bibinfo {author} {\bibfnamefont {F.~N.}\ \bibnamefont {\"Unal}},
  \bibinfo {author} {\bibfnamefont {A.}~\bibnamefont {Eckardt}}, \bibinfo
  {author} {\bibfnamefont {M.}~\bibnamefont {{Di Liberto}}}, \bibinfo {author}
  {\bibfnamefont {N.}~\bibnamefont {Goldman}}, \bibinfo {author} {\bibfnamefont
  {I.}~\bibnamefont {Bloch}}, \ and\ \bibinfo {author} {\bibfnamefont
  {M.}~\bibnamefont {Aidelsburger}},\ }\href {\doibase
  10.1038/s41567-020-0949-y} {\bibfield  {journal} {\bibinfo  {journal} {Nat.
  Phys.}\ } (\bibinfo {year} {2020}),\ 10.1038/s41567-020-0949-y}\BibitemShut
  {NoStop}%
\bibitem [{\citenamefont {Joglekar}\ \emph {et~al.}(2014)\citenamefont
  {Joglekar}, \citenamefont {Marathe}, \citenamefont {Durganandini},\ and\
  \citenamefont {Pathak}}]{Joglekar2014}%
  \BibitemOpen
  \bibfield  {author} {\bibinfo {author} {\bibfnamefont {Y.~N.}\ \bibnamefont
  {Joglekar}}, \bibinfo {author} {\bibfnamefont {R.}~\bibnamefont {Marathe}},
  \bibinfo {author} {\bibfnamefont {P.}~\bibnamefont {Durganandini}}, \ and\
  \bibinfo {author} {\bibfnamefont {R.~K.}\ \bibnamefont {Pathak}},\ }\href
  {\doibase 10.1103/physreva.90.040101} {\bibfield  {journal} {\bibinfo
  {journal} {Phys. Rev. A}\ }\textbf {\bibinfo {volume} {90}},\ \bibinfo
  {pages} {040101} (\bibinfo {year} {2014})}\BibitemShut {NoStop}%
\bibitem [{\citenamefont {Lee}\ and\ \citenamefont {Joglekar}(2015)}]{Lee2015}%
  \BibitemOpen
  \bibfield  {author} {\bibinfo {author} {\bibfnamefont {T.~E.}\ \bibnamefont
  {Lee}}\ and\ \bibinfo {author} {\bibfnamefont {Y.~N.}\ \bibnamefont
  {Joglekar}},\ }\href {\doibase 10.1103/physreva.92.042103} {\bibfield
  {journal} {\bibinfo  {journal} {Phys. Rev. A}\ }\textbf {\bibinfo {volume}
  {92}},\ \bibinfo {pages} {042103} (\bibinfo {year} {2015})}\BibitemShut
  {NoStop}%
\bibitem [{\citenamefont {Gong}\ and\ \citenamefont {hai
  Wang}(2015)}]{Gong2015}%
  \BibitemOpen
  \bibfield  {author} {\bibinfo {author} {\bibfnamefont {J.}~\bibnamefont
  {Gong}}\ and\ \bibinfo {author} {\bibfnamefont {Q.}~\bibnamefont {hai
  Wang}},\ }\href {\doibase 10.1103/physreva.91.042135} {\bibfield  {journal}
  {\bibinfo  {journal} {Phys. Rev. A}\ }\textbf {\bibinfo {volume} {91}},\
  \bibinfo {pages} {042135} (\bibinfo {year} {2015})}\BibitemShut {NoStop}%
\bibitem [{\citenamefont {Longhi}(2017{\natexlab{a}})}]{Longhi2017a}%
  \BibitemOpen
  \bibfield  {author} {\bibinfo {author} {\bibfnamefont {S.}~\bibnamefont
  {Longhi}},\ }\href {\doibase 10.1209/0295-5075/117/10005} {\bibfield
  {journal} {\bibinfo  {journal} {Europhys. Lett.}\ }\textbf {\bibinfo {volume}
  {117}},\ \bibinfo {pages} {10005} (\bibinfo {year}
  {2017}{\natexlab{a}})}\BibitemShut {NoStop}%
\bibitem [{\citenamefont {Longhi}(2017{\natexlab{b}})}]{Longhi2017b}%
  \BibitemOpen
  \bibfield  {author} {\bibinfo {author} {\bibfnamefont {S.}~\bibnamefont
  {Longhi}},\ }\href {\doibase 10.1088/1751-8121/aa931f} {\bibfield  {journal}
  {\bibinfo  {journal} {J. Phys. A}\ }\textbf {\bibinfo {volume} {50}},\
  \bibinfo {pages} {505201} (\bibinfo {year} {2017}{\natexlab{b}})}\BibitemShut
  {NoStop}%
\bibitem [{\citenamefont {Zhou}\ and\ \citenamefont {Gong}(2018)}]{Zhou2018}%
  \BibitemOpen
  \bibfield  {author} {\bibinfo {author} {\bibfnamefont {L.}~\bibnamefont
  {Zhou}}\ and\ \bibinfo {author} {\bibfnamefont {J.}~\bibnamefont {Gong}},\
  }\href {\doibase 10.1103/physrevb.98.205417} {\bibfield  {journal} {\bibinfo
  {journal} {Phys. Rev. B}\ }\textbf {\bibinfo {volume} {98}},\ \bibinfo
  {pages} {205417} (\bibinfo {year} {2018})}\BibitemShut {NoStop}%
\bibitem [{\citenamefont {Zhou}(2019)}]{Zhou2019}%
  \BibitemOpen
  \bibfield  {author} {\bibinfo {author} {\bibfnamefont {L.}~\bibnamefont
  {Zhou}},\ }\href {\doibase 10.1103/physrevb.100.184314} {\bibfield  {journal}
  {\bibinfo  {journal} {Phys. Rev. B}\ }\textbf {\bibinfo {volume} {100}},\
  \bibinfo {pages} {184314} (\bibinfo {year} {2019})}\BibitemShut {NoStop}%
\bibitem [{\citenamefont {Li}\ \emph {et~al.}(2019{\natexlab{a}})\citenamefont
  {Li}, \citenamefont {Ni}, \citenamefont {Weiner}, \citenamefont {Al\`u},\
  and\ \citenamefont {Khanikaev}}]{Li2019b}%
  \BibitemOpen
  \bibfield  {author} {\bibinfo {author} {\bibfnamefont {M.}~\bibnamefont
  {Li}}, \bibinfo {author} {\bibfnamefont {X.}~\bibnamefont {Ni}}, \bibinfo
  {author} {\bibfnamefont {M.}~\bibnamefont {Weiner}}, \bibinfo {author}
  {\bibfnamefont {A.}~\bibnamefont {Al\`u}}, \ and\ \bibinfo {author}
  {\bibfnamefont {A.~B.}\ \bibnamefont {Khanikaev}},\ }\href {\doibase
  10.1103/physrevb.100.045423} {\bibfield  {journal} {\bibinfo  {journal}
  {Phys. Rev. B}\ }\textbf {\bibinfo {volume} {100}},\ \bibinfo {pages}
  {045423} (\bibinfo {year} {2019}{\natexlab{a}})}\BibitemShut {NoStop}%
\bibitem [{\citenamefont {Zhang}\ and\ \citenamefont {Gong}(2020)}]{Zhang2020}%
  \BibitemOpen
  \bibfield  {author} {\bibinfo {author} {\bibfnamefont {X.}~\bibnamefont
  {Zhang}}\ and\ \bibinfo {author} {\bibfnamefont {J.}~\bibnamefont {Gong}},\
  }\href {\doibase 10.1103/physrevb.101.045415} {\bibfield  {journal} {\bibinfo
   {journal} {Phys. Rev. B}\ }\textbf {\bibinfo {volume} {101}},\ \bibinfo
  {pages} {045415} (\bibinfo {year} {2020})}\BibitemShut {NoStop}%
\bibitem [{\citenamefont {Wu}\ and\ \citenamefont {An}(2020)}]{Wu2020}%
  \BibitemOpen
  \bibfield  {author} {\bibinfo {author} {\bibfnamefont {H.}~\bibnamefont
  {Wu}}\ and\ \bibinfo {author} {\bibfnamefont {J.-H.}\ \bibnamefont {An}},\
  }\href {\doibase 10.1103/physrevb.102.041119} {\bibfield  {journal} {\bibinfo
   {journal} {Phys. Rev. B}\ }\textbf {\bibinfo {volume} {102}},\ \bibinfo
  {pages} {041119} (\bibinfo {year} {2020})}\BibitemShut {NoStop}%
\bibitem [{\citenamefont {Yuce}(2015{\natexlab{a}})}]{Yuce2015a}%
  \BibitemOpen
  \bibfield  {author} {\bibinfo {author} {\bibfnamefont {C.}~\bibnamefont
  {Yuce}},\ }\href {\doibase 10.1140/epjd/e2014-50652-x} {\bibfield  {journal}
  {\bibinfo  {journal} {Eur. Phys. J. D}\ }\textbf {\bibinfo {volume} {69}},\
  \bibinfo {pages} {11} (\bibinfo {year} {2015}{\natexlab{a}})}\BibitemShut
  {NoStop}%
\bibitem [{\citenamefont {Yuce}(2015{\natexlab{b}})}]{Yuce2015b}%
  \BibitemOpen
  \bibfield  {author} {\bibinfo {author} {\bibfnamefont {C.}~\bibnamefont
  {Yuce}},\ }\href {\doibase 10.1140/epjd/e2015-60220-7} {\bibfield  {journal}
  {\bibinfo  {journal} {Eur. Phys. J. D}\ }\textbf {\bibinfo {volume} {69}},\
  \bibinfo {pages} {184} (\bibinfo {year} {2015}{\natexlab{b}})}\BibitemShut
  {NoStop}%
\bibitem [{\citenamefont {Maamache}\ \emph {et~al.}(2017)\citenamefont
  {Maamache}, \citenamefont {Lamri},\ and\ \citenamefont
  {Cherbal}}]{Maamache2017}%
  \BibitemOpen
  \bibfield  {author} {\bibinfo {author} {\bibfnamefont {M.}~\bibnamefont
  {Maamache}}, \bibinfo {author} {\bibfnamefont {S.}~\bibnamefont {Lamri}}, \
  and\ \bibinfo {author} {\bibfnamefont {O.}~\bibnamefont {Cherbal}},\ }\href
  {\doibase 10.1016/j.aop.2017.01.005} {\bibfield  {journal} {\bibinfo
  {journal} {Ann. Phys.}\ }\textbf {\bibinfo {volume} {378}},\ \bibinfo {pages}
  {150} (\bibinfo {year} {2017})}\BibitemShut {NoStop}%
\bibitem [{\citenamefont {Turker}\ \emph {et~al.}(2018)\citenamefont {Turker},
  \citenamefont {Tombuloglu},\ and\ \citenamefont {Yuce}}]{Turker2018}%
  \BibitemOpen
  \bibfield  {author} {\bibinfo {author} {\bibfnamefont {Z.}~\bibnamefont
  {Turker}}, \bibinfo {author} {\bibfnamefont {S.}~\bibnamefont {Tombuloglu}},
  \ and\ \bibinfo {author} {\bibfnamefont {C.}~\bibnamefont {Yuce}},\ }\href
  {\doibase 10.1016/j.physleta.2018.05.015} {\bibfield  {journal} {\bibinfo
  {journal} {Phys. Lett. A}\ }\textbf {\bibinfo {volume} {382}},\ \bibinfo
  {pages} {2013} (\bibinfo {year} {2018})}\BibitemShut {NoStop}%
\bibitem [{\citenamefont {Harter}\ and\ \citenamefont
  {Hatano}(2020)}]{Harter2020}%
  \BibitemOpen
  \bibfield  {author} {\bibinfo {author} {\bibfnamefont {A.~K.}\ \bibnamefont
  {Harter}}\ and\ \bibinfo {author} {\bibfnamefont {N.}~\bibnamefont
  {Hatano}},\ }\href@noop {} {\enquote {\bibinfo {title} {Real edge modes in a
  {Floquet}-modulated {$\mathcal{PT}$}-symmetric {SSH} model},}\ } (\bibinfo
  {year} {2020}),\ \Eprint {http://arxiv.org/abs/2006.16890} {arXiv:2006.16890
  [quant-ph]} \BibitemShut {NoStop}%
\bibitem [{\citenamefont {Li}\ \emph {et~al.}(2020)\citenamefont {Li},
  \citenamefont {Wang}, \citenamefont {Luo}, \citenamefont {Vemuri},\ and\
  \citenamefont {Joglekar}}]{Li2020}%
  \BibitemOpen
  \bibfield  {author} {\bibinfo {author} {\bibfnamefont {J.}~\bibnamefont
  {Li}}, \bibinfo {author} {\bibfnamefont {T.}~\bibnamefont {Wang}}, \bibinfo
  {author} {\bibfnamefont {L.}~\bibnamefont {Luo}}, \bibinfo {author}
  {\bibfnamefont {S.}~\bibnamefont {Vemuri}}, \ and\ \bibinfo {author}
  {\bibfnamefont {Y.~N.}\ \bibnamefont {Joglekar}},\ }\href@noop {} {\enquote
  {\bibinfo {title} {Unification of quantum {Zeno}-anti {Zeno} effects and
  parity-time symmetry breaking transitions},}\ } (\bibinfo {year} {2020}),\
  \Eprint {http://arxiv.org/abs/2004.01364} {arXiv:2004.01364 [quant-ph]}
  \BibitemShut {NoStop}%
\bibitem [{\citenamefont {Duan}\ \emph {et~al.}(2020)\citenamefont {Duan},
  \citenamefont {Wang},\ and\ \citenamefont {Chen}}]{Duan2020}%
  \BibitemOpen
  \bibfield  {author} {\bibinfo {author} {\bibfnamefont {L.}~\bibnamefont
  {Duan}}, \bibinfo {author} {\bibfnamefont {Y.-Z.}\ \bibnamefont {Wang}}, \
  and\ \bibinfo {author} {\bibfnamefont {Q.-H.}\ \bibnamefont {Chen}},\ }\href
  {\doibase 10.1088/0256-307x/37/8/081101} {\bibfield  {journal} {\bibinfo
  {journal} {Chin. Phys. Lett.}\ }\textbf {\bibinfo {volume} {37}},\ \bibinfo
  {pages} {081101} (\bibinfo {year} {2020})}\BibitemShut {NoStop}%
\bibitem [{\citenamefont {Mochizuki}\ \emph {et~al.}(2020)\citenamefont
  {Mochizuki}, \citenamefont {Kim}, \citenamefont {Kawakami},\ and\
  \citenamefont {Obuse}}]{Mochizuki2020}%
  \BibitemOpen
  \bibfield  {author} {\bibinfo {author} {\bibfnamefont {K.}~\bibnamefont
  {Mochizuki}}, \bibinfo {author} {\bibfnamefont {D.}~\bibnamefont {Kim}},
  \bibinfo {author} {\bibfnamefont {N.}~\bibnamefont {Kawakami}}, \ and\
  \bibinfo {author} {\bibfnamefont {H.}~\bibnamefont {Obuse}},\ }\href
  {\doibase 10.1103/physreva.102.062202} {\bibfield  {journal} {\bibinfo
  {journal} {Phys. Rev. A}\ }\textbf {\bibinfo {volume} {102}},\ \bibinfo
  {pages} {062202} (\bibinfo {year} {2020})}\BibitemShut {NoStop}%
\bibitem [{\citenamefont {Chitsazi}\ \emph {et~al.}(2017)\citenamefont
  {Chitsazi}, \citenamefont {Li}, \citenamefont {Ellis},\ and\ \citenamefont
  {Kottos}}]{Chitsazi2017}%
  \BibitemOpen
  \bibfield  {author} {\bibinfo {author} {\bibfnamefont {M.}~\bibnamefont
  {Chitsazi}}, \bibinfo {author} {\bibfnamefont {H.}~\bibnamefont {Li}},
  \bibinfo {author} {\bibfnamefont {F.~M.}\ \bibnamefont {Ellis}}, \ and\
  \bibinfo {author} {\bibfnamefont {T.}~\bibnamefont {Kottos}},\ }\href
  {\doibase 10.1103/physrevlett.119.093901} {\bibfield  {journal} {\bibinfo
  {journal} {Phys. Rev. Lett.}\ }\textbf {\bibinfo {volume} {119}},\ \bibinfo
  {pages} {093901} (\bibinfo {year} {2017})}\BibitemShut {NoStop}%
\bibitem [{\citenamefont {Li}\ \emph {et~al.}(2019{\natexlab{b}})\citenamefont
  {Li}, \citenamefont {Harter}, \citenamefont {Liu}, \citenamefont {de~Melo},
  \citenamefont {Joglekar},\ and\ \citenamefont {Luo}}]{Li2019a}%
  \BibitemOpen
  \bibfield  {author} {\bibinfo {author} {\bibfnamefont {J.}~\bibnamefont
  {Li}}, \bibinfo {author} {\bibfnamefont {A.~K.}\ \bibnamefont {Harter}},
  \bibinfo {author} {\bibfnamefont {J.}~\bibnamefont {Liu}}, \bibinfo {author}
  {\bibfnamefont {L.}~\bibnamefont {de~Melo}}, \bibinfo {author} {\bibfnamefont
  {Y.~N.}\ \bibnamefont {Joglekar}}, \ and\ \bibinfo {author} {\bibfnamefont
  {L.}~\bibnamefont {Luo}},\ }\href {\doibase 10.1038/s41467-019-08596-1}
  {\bibfield  {journal} {\bibinfo  {journal} {Nat. Commun.}\ }\textbf {\bibinfo
  {volume} {10}},\ \bibinfo {pages} {855} (\bibinfo {year}
  {2019}{\natexlab{b}})}\BibitemShut {NoStop}%
\bibitem [{\citenamefont {de~J.~Le\'on-Montiel}\ \emph
  {et~al.}(2018)\citenamefont {de~J.~Le\'on-Montiel}, \citenamefont
  {Quiroz-Ju\'arez}, \citenamefont {Dom\'inguez-Ju\'arez}, \citenamefont
  {Quintero-Torres}, \citenamefont {Arag\'on}, \citenamefont {Harter},\ and\
  \citenamefont {Joglekar}}]{LeonMontiel2018}%
  \BibitemOpen
  \bibfield  {author} {\bibinfo {author} {\bibfnamefont {R.}~\bibnamefont
  {de~J.~Le\'on-Montiel}}, \bibinfo {author} {\bibfnamefont {M.~A.}\
  \bibnamefont {Quiroz-Ju\'arez}}, \bibinfo {author} {\bibfnamefont {J.~L.}\
  \bibnamefont {Dom\'inguez-Ju\'arez}}, \bibinfo {author} {\bibfnamefont
  {R.}~\bibnamefont {Quintero-Torres}}, \bibinfo {author} {\bibfnamefont
  {J.~L.}\ \bibnamefont {Arag\'on}}, \bibinfo {author} {\bibfnamefont {A.~K.}\
  \bibnamefont {Harter}}, \ and\ \bibinfo {author} {\bibfnamefont {Y.~N.}\
  \bibnamefont {Joglekar}},\ }\href {\doibase 10.1038/s42005-018-0087-3}
  {\bibfield  {journal} {\bibinfo  {journal} {Commun. Phys.}\ }\textbf
  {\bibinfo {volume} {1}},\ \bibinfo {pages} {88} (\bibinfo {year}
  {2018})}\BibitemShut {NoStop}%
\bibitem [{\citenamefont {Joglekar}\ and\ \citenamefont
  {Harter}(2018)}]{Joglekar2018}%
  \BibitemOpen
  \bibfield  {author} {\bibinfo {author} {\bibfnamefont {Y.~N.}\ \bibnamefont
  {Joglekar}}\ and\ \bibinfo {author} {\bibfnamefont {A.~K.}\ \bibnamefont
  {Harter}},\ }\href {\doibase 10.1364/prj.6.000a51} {\bibfield  {journal}
  {\bibinfo  {journal} {Photon. Res.}\ }\textbf {\bibinfo {volume} {6}},\
  \bibinfo {pages} {A51} (\bibinfo {year} {2018})}\BibitemShut {NoStop}%
\bibitem [{\citenamefont {Caves}(1982)}]{Caves1982}%
  \BibitemOpen
  \bibfield  {author} {\bibinfo {author} {\bibfnamefont {C.~M.}\ \bibnamefont
  {Caves}},\ }\href {\doibase 10.1103/PhysRevD.26.1817} {\bibfield  {journal}
  {\bibinfo  {journal} {Phys. Rev. D}\ }\textbf {\bibinfo {volume} {26}},\
  \bibinfo {pages} {1817} (\bibinfo {year} {1982})}\BibitemShut {NoStop}%
\bibitem [{\citenamefont {Scheel}\ and\ \citenamefont
  {Szameit}(2018)}]{Scheel2018}%
  \BibitemOpen
  \bibfield  {author} {\bibinfo {author} {\bibfnamefont {S.}~\bibnamefont
  {Scheel}}\ and\ \bibinfo {author} {\bibfnamefont {A.}~\bibnamefont
  {Szameit}},\ }\href {\doibase 10.1209/0295-5075/122/34001} {\bibfield
  {journal} {\bibinfo  {journal} {Europhys. Lett.}\ }\textbf {\bibinfo {volume}
  {122}},\ \bibinfo {pages} {34001} (\bibinfo {year} {2018})}\BibitemShut
  {NoStop}%
\end{thebibliography}%

\end{document}